\documentclass[aps, prd, amsmath, floats, floatfix, twocolumn,
superscriptaddress, nofootinbib, showpacs]{revtex4}
\usepackage{graphicx}
\usepackage{epsfig}
\usepackage{color}
\usepackage{soul}
\usepackage{url}
\usepackage{bm}         
\usepackage{times}

\newcommand{\beq}{\begin{equation}}
\newcommand{\eeq}{\end{equation}}
\newcommand{\beqn}{\begin{eqnarray}}
\newcommand{\eeqn}{\end{eqnarray}}

\newcommand{\lo}{\mathrel{\raise.3ex\hbox{$<$}\mkern-14mu
    \lower0.6ex\hbox{$\sim$}}}
\newcommand{\go}{\mathrel{\raise.3ex\hbox{$>$}\mkern-14mu
    \lower0.6ex\hbox{$\sim$}}}

\usepackage{color}

\newcommand{\Caltech}{\affiliation{Theoretical Astrophysics 350-17,
    California Institute of Technology, Pasadena, California 91125, USA}}
\newcommand{\Cornell}{\affiliation{Center for Radiophysics and Space
    Research, Cornell University, Ithaca, New York, 14853, USA}}
\newcommand{\WSU}{\affiliation{Department of Physics \& Astronomy,
	Washington State University, Pullman, Washington 99164, USA}}
\newcommand{\CITA}{\affiliation{Canadian Institute for Theoretical 
    Astrophysics, University of Toronto, Toronto, Ontario M5S 3H8, Canada}}
\newcommand{\UofT}{\affiliation{Department of Astronomy \& Astrophysics,
    University of Toronto, Toronto, Ontario, M5S 3H5, Canada}}

\usepackage{graphicx}
\usepackage{dcolumn}
\usepackage{bm}
\usepackage{epsf}

\begin{document}

\title{Black-hole--neutron-star mergers at realistic mass ratios: Equation of state and spin orientation effects}

\author{Francois Foucart} \CITA%
\author{M. Brett Deaton} \WSU %
\author{Matthew D. Duez} \WSU %
\author{Lawrence E. Kidder} \Cornell %
\author{Ilana MacDonald} \CITA \UofT
\author{Christian D. Ott} \Caltech 
\author{Harald P. Pfeiffer} \CITA
\author{Mark A. Scheel} \Caltech %
\author{Bela Szilagyi} \Caltech %
\author{Saul A. Teukolsky} \Cornell

\begin{abstract}

Black-hole--neutron-star mergers resulting in the disruption of the neutron star and the formation of an accretion disk and/or
the ejection of unbound material are prime candidates for the joint detection of gravitational-wave and electromagnetic signals when
the next generation of gravitational-wave detectors comes online. However, the disruption of the neutron star and the properties of the 
post-merger remnant are very sensitive to the parameters of the binary (mass ratio, black hole spin, neutron star radius). In this paper, 
we study the impact of the radius of the neutron star and the alignment of the black hole spin on black-hole--neutron-star mergers within the range
of mass ratio currently deemed most likely for field binaries ($M_{\rm BH}\sim 7 M_{\rm NS}$) and for black hole spins large enough for
the neutron star to disrupt ($J_{\rm BH}/M_{\rm BH}^2=0.9$). We find that: 
(i) In this regime, the merger is particularly sensitive to the radius of the neutron star, 
with remnant masses varying from $0.3M_{\rm NS}$ to $0.1M_{\rm NS}$ for changes of only $2\,{\rm km}$ in the NS radius; 
(ii) $0.01M_\odot-0.05M_\odot$ of unbound material can be ejected with kinetic energy $\go 10^{51}\,{\rm ergs}$,
a significant increase compared to low mass ratio, low spin binaries. This ejecta could power detectable post-merger optical and radio afterglows.  
(iii) Only a small fraction of the Advanced LIGO events in this parameter range have gravitational-wave signals which could offer
constraints on the equation of state of the neutron star (at best $\sim 3\%$ of the events for a single detector at design sensitivity).
(iv) A misaligned black hole spin works against disk formation, with less neutron star material remaining outside of the black hole after merger,
and a larger fraction of that material remaining in the tidal tail instead of the forming accretion disk. 
(v) Large kicks $v_{\rm kick}\go 300\,{\rm km/s}$ can be given to the final black hole
as a result of a precessing BHNS merger, when the disruption of the neutron star occurs just outside or within the innermost stable
spherical orbit.

\end{abstract}

\pacs{04.25.dg, 04.30.-w, 04.40.Dg, 47.75.+f}

\maketitle

\section{Introduction}
\label{intro}

The next generation of ground-based gravitational-wave detectors (Advanced LIGO, Advanced Virgo and KAGRA~\cite{LIGO,VIRGO,2012CQGra..29l4007S}) 
will progressively begin taking data over the next decade, opening an entirely
new way to observe the universe. One of the sources of gravitational waves that they will detect are compact binary coalescences: binary black holes (BBH), black-hole--neutron-star (BHNS) and
binary neutron star (BNS) systems~\cite{2010CQGra..27q3001A}. 
In the presence of a neutron star, these gravitational-wave signals could be accompanied by electromagnetic emissions, which would provide better sky 
localization and additional information about the characteristics and the environment of the binary. The most energetic, and most often discussed potential counterparts are short gamma-ray
bursts (SGRBs, see e.g.~\cite{2007NJPh....9...17L}), while other possibilities include the x-ray and optical afterglows of SGRBs, 
optical transients due to the radioactive decay of neutron-rich unbound material, and radio emission from that ejecta as it decelerates 
in the interstellar medium (see~\cite{2012ApJ...746...48M} for more details on EM signals emitted by compact binary mergers,
and~\cite{2012arXiv1210.6362N} for their detectability). 
The ejection of a small amount of
material at ultrarelativistic speeds as a result of a shock in the region in which two neutron stars first get into contact was also recently proposed as a potential outcome of 
BNS mergers~\cite{2012arXiv1209.5747K}.
Finally, pre-merger electromagnetic transients are also a possibility, and could for example be due to the breaking of the neutron star crust~\cite{2012PhRvL.108a1102T}.

Which of these effects occur in practice depends strongly on the parameters of the binary. The exact conditions leading to the emission of a SGRB are not known, and will
depend on the physical process responsible for these bursts --- most likely either the extraction of black hole rotational energy by the magnetic field of the accretion disk (Blandford-Znajek 
effect~\cite{1977MNRAS.179..433B}), magnetically-driven outflows in the accretion disk (Blandford-Payne effect~\cite{1982MNRAS.199..883B}),
or the production of ultrarelativistic $e^+e^-$ pairs from the annihilation of $\nu \tilde \nu$ pairs, themselves produced by a hot accretion disk surrounding the remnant 
black hole~\cite{Lee:2005se,2007NJPh....9...17L}. In all cases, the presence of both an accretion disk 
and a baryon-poor region in which a relativistic jet can be produced
appears to be a prerequisite. The amount of matter required in the disk depends on the efficiency of the jet production mechanism and the energy of the burst, 
with estimates spanning multiple orders of magnitudes ($10^{-4}M_\odot-0.5 M_\odot$)~\cite{2005ApJ...630L.165L,2012arXiv1210.8152G}.
The formation of an accretion disk is a natural result of the merger of BNS systems. For BHNS binaries, however, many initial conditions lead to the direct plunge
of the neutron star into the black hole, before the tidal field of the hole can disrupt the neutron star and cause the formation of an accretion disk. 
Relatively massive black holes ($M_{\rm BH} \go 7M_\odot$) are favored both by current population synthesis models~\cite{2008ApJ...682..474B,2010ApJ...715L.138B}
and by the distribution of black hole masses measured in galactic X-ray binaries~\cite{2010ApJ...725.1918O}.
In that regime, tidal disruption only occurs for the most rapidly spinning black holes~\cite{2012PhRvD..85d4015F}. 

The ejection of enough unbound material to power detectable electromagnetic transients is not a certainty either. 
General relativistic simulations of BNS systems have found 
that only a small amount of mass is unbound by the merger ($M_{ej}=10^{-4}M_\odot-10^{-2}M_\odot$)~\cite{2012arXiv1212.0905H}, 
while Newtonian smoothed particle hydrodynamics (SPH) simulations have more optimistic 
predictions ($M_{ej}\go 10^{-2}M_\odot$)~\cite{2012arXiv1206.2379K}.
BHNS systems have more 
asymmetric mass ratios, and are thus generally more favorable for the ejection of neutron-rich material during the disruption of the neutron star. However, this requires the neutron star to be 
disrupted in the first place. Numerical simulations have shown that this will only be the case if the mass of the black hole is lower than what population synthesis models predict, or if the spin of the
black hole is high. 
The amount of material ejected in such mergers remains very uncertain. SPH results predict ejected masses of up to $\sim 0.1M_\odot$~\cite{2005ApJ...634.1202R}. 
General relativistic simulations for low mass, low spin black holes find little ejected material ($<0.01M_\odot$)~\cite{Duez:2010a}, but estimates for
rapidly spinning black holes have not been offered yet. We
show in this paper that more massive outflows ($0.01M_\odot-0.05M_\odot$) are likely for high black hole spins.  
Finally, matter could be ejected due to magnetically-driven~\cite{2011ApJ...734L..36S} or neutrino-driven~\cite{2009ApJ...690.1681D} winds in the disk.

An important limitation of existing general relativistic simulations of BHNS mergers is the lack of coverage of the range of black hole masses
deemed most likely astrophysically. The only simulations
considering mass ratios $q=M_{\rm BH}/M_{\rm NS}>5$ 
have shown that, even for very large neutron stars ($R_{\rm NS}\sim 14.4\,{\rm km}$) and aligned black hole spins, dimensionless black hole 
spins $\chi_{\rm BH}=J/M_{\rm BH}^2 \go 0.7$ are required for tidal disruption to occur~\cite{2012PhRvD..85d4015F}.
Simulations for more symmetric mass ratios have however already shown that 
smaller neutron star 
radii~\cite{Duez:2010a,PhysRevD.84.064018} or misaligned black hole spins~\cite{2011PhRvD..83b4005F} are likely to make tidal disruption harder. 
In this paper, we begin a more quantitative exploration of these effects. We consider BHNS binaries at mass ratio $q=7$ and with large
black hole spins $\chi_{\rm BH}=0.9$, and vary the radius of the neutron star and the orientation of the black hole spin. We currently limit ourselves to the most basic physical effects which
will influence the dynamics of the merger: we treat gravity in a fully general relativistic framework, but use a simple $\Gamma$-law equation of state to describe the neutron star matter, and ignore
the effects of magnetic fields or neutrino radiation. More realistic equations of state are likely to influence tidal disruption, but to first order the compactness of the neutron star is expected to capture
the main physical effects during merger~\cite{Duez:2010a}. 
Magnetic fields are important for the evolution of any post-merger accretion disk, as these disks are unstable 
to the magnetorotational instability (MRI), but very large internal fields are necessary for magnetic effects to influence the disruption of the neutron 
star~\cite{2010PhRvL.105k1101C,2012PhRvD..85f4029E}. Finally, while neutrinos are the main source
of cooling in the disk, and are thus crucial to their evolution over a cooling timescale $\tau_{\nu}$ ($\tau_{\nu}\sim 0.1s$ for the relatively
dense disks considered by Lee et al.~\cite{Lee:2005se}, and could be significantly smaller for the lower density disks observed at the end of our simulations),
neutrino emission will be negligible before merger (neutron stars are expected to be extremely cool, with $T \ll 1\,{\rm MeV}$).
This was confirmed numerically in the case of BNS~\cite{2012CQGra..29l4003K}.
Our simulations will thus capture the physical effects which are important for the evolution of BHNS systems in the last few orbits before their merger, and during
the merger itself. They are however not suitable for the long term evolution of the post-merger remnant.

Gravitational waveforms from BHNS and BNS mergers are of particular interest for the constraints that they might offer on the unknown equation of state of
the neutron star. Numerical simulations indicate that from the last few orbits of a BNS merger occurring at $100\,{\rm Mpc}$, constraints of $\sim 1\,{\rm km}$ could be obtained
on the radius of the neutron star~\cite{2009PhRvD..79l4033R}. Damour et al.~\cite{2012PhRvD..85l3007D} have shown using Effective One Body waveforms that equations of state effects in the late inspiral could
be measured for events of moderate signal-to-noise ratio ($\rho\sim 16$). Accurate numerical simulations are however necessary to calibrate such models at high frequency.
Numerical results by Bernuzzi et al.~\cite{2012PhRvD..86d4030B} have confirmed the predictions of Damour et al.~\cite{2012PhRvD..85l3007D} regarding the detectability of these equations of state effects, but existing
simulations are not accurate enough to model the waveform with the accuracy required to take full advantage of all of the information that will be available in waveforms detected by Advanced LIGO.
For BHNS mergers of nonspinning black holes, tidal effects during the inspiral are too small to be detected directly by Advanced LIGO~\cite{2010PhRvD..81l3016H,2012PhRvD..85l3007D}. 
But the cutoff in the gravitational-wave spectrum occurring when the neutron star is disrupted by the tidal field of the black hole can be. 
Semi-analytical models have been developed to attempt to extract that information~\cite{2000PhRvL..84.3519V,ferrariBHNS:09}. Numerical simulations mapping the cutoff frequency
across the relevant parameter space are however necessary to better calibrate them. 

At low mass ratios ($q=2-3$) and for nonspinning black holes, 
Lackey et al.~\cite{2012PhRvD..85d4061L} have shown from numerical simulations that the combined effects of the tidal interactions during the inspiral 
and of the high-frequency cutoff of the signal would allow Advanced LIGO to detect variations of $\sim 10\%-40\%$ in the radius of the neutron star for a favorable event at $\sim 100\,{\rm Mpc}$. 
This is thus slightly inferior to what can be done for binary neutron star systems located at the same distance from the observer
(and, within a fixed volume, we expect many more BNS mergers than BHNS mergers). At higher mass ratio, tidal effects during the inspiral further decrease~\cite{2010PhRvD..81l3016H}.
 On the other hand, due to the higher total mass of the system, the amplitude of the signal will be larger. Tidal disruption will also occur at lower frequency, 
and thus in a more favorable region of the LIGO noise curve. The spin of the black hole can also affect how much tidal distortion occurs during the inspiral 
(as the neutron star can get closer to a spinning BH before reaching its ISCO, tidal effects can be stronger). 
How these competing effects will affect our ability to measure the properties of neutron stars in BHNS binaries, 
or even to differentiate BHNS systems from BBH binaries, remains an important open question.

In this paper, we study the influence of the radius of the neutron star and of the orientation of the black hole spin on the dynamics of the merger of BHNS binaries around the black hole mass $M_{\rm BH}\sim 10M_\odot$ currently favored by population synthesis models, focusing on tidal effects during the inspiral, on the initial characteristics of the post-merger remnant (accretion disk and tidal tail formation), and on the properties
of the emitted gravitational-wave signal --- and in particular the effects of the neutron star radius on that signal, and the conditions under which such effects might be detected by the 
next generation of gravitational-wave experiments. We stop the simulations $5\,{\rm ms}$ after merger, as
neglected microphysical effects are likely to become important at later times. We will begin by describing briefly the numerical setup used for our simulations, as well as modifications to the code since the publication
of~\cite{2012PhRvD..85d4015F} (Sec.~\ref{sec:ns}). 
We will then detail the initial configurations evolved (Sec.~\ref{sec:id}), and estimate the accuracy of the results (Sec.~\ref{sec:acc}). Finally, the main physical results are presented in 
Sec.~\ref{sec:nr}.

\section{Numerical Setup}
\label{sec:ns}

\begin{figure}
\includegraphics[width=0.95\columnwidth]{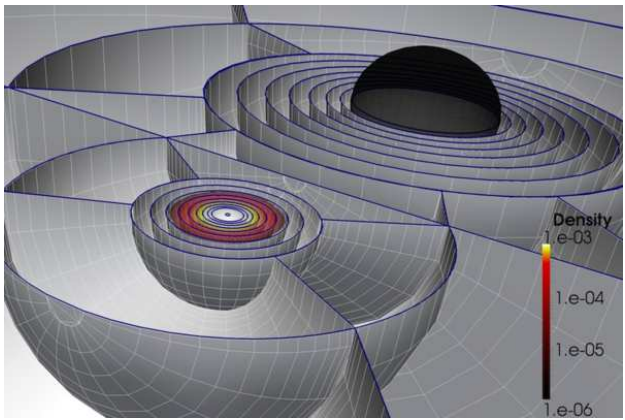}
\caption{Numerical grid before disruption of the neutron star, below the equatorial plane of the binary. The black hole is the excised region on the right (black sphere), while we superpose 
a linear color scale for the baryon density $\rho_0$. The subdomains on the outer edge of the plot connect to spherical shells covering the wave region.}
\label{fig:gridinsp}
\end{figure}

The numerical simulations presented here are performed with the SpEC code~\cite{SpEC}, which evolves Einstein's equations of general relativity coupled
to the relativistic hydrodynamics equations (see Appendix~\ref{app:eqns} for details). 
Einstein's equations are solved using pseudospectral methods, in the generalized harmonics formulation~\cite{Lindblom:2006}, and excising
the black hole interior.
The numerical grid on which Einstein's equations are solved consists at first of 8 spherical shells immediately around the black hole, 8 spherical shells and 1 inner ball in the region close to the neutron star,
24 spherical shells covering the far-field region, and a set of 13 distorted cylindrical shells and filled cylinders connecting them (see Fig.~\ref{fig:gridinsp}). All subdomains
are touching but not overlapping. The low, medium and high resolution runs correspond to a total number of points $N=57^3,64^3,72^3$. 

\begin{figure}
\includegraphics[width=0.95\columnwidth]{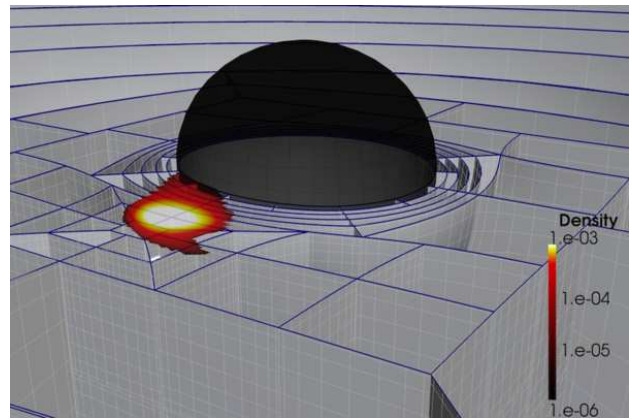}
\caption{Same as Fig.~\ref{fig:gridinsp}, but during the disruption of the neutron star.}
\label{fig:gridmerger}
\end{figure}

Once the neutron star disrupts, we cannot take advantage of an approximate spherical symmetry
around the neutron star, and have to modify the pseudospectral grid. The shells around the black hole are replaced by a set of 264 distorted cubes,
in order to allow for high angular resolution as the neutron star falls into the hole. The wave zone is still covered by 24 spherical shells, while the region around the neutron
star and the near field region are covered by non-overlapping distorted cubes (see Fig.~\ref{fig:gridmerger}). The resolution is chosen adaptively, by requiring that the
relative truncation error for each set of basis functions (measured from the coefficients of the spectral expansion of the evolved variables) is $(0.5,0.7,1.0) \times 10^{-4}$
for our 3 resolutions. The actual number of grid points thus vary during the merger, peaking as the neutron star accretes onto the black hole. At medium resolution, 
we have $N\sim 70^3 - 110^3$.

The relativistic hydrodynamics equations are solved on a separate finite difference grid~\cite{Duez:2008rb}. 
The grid covers only the region in which matter is present, and expands/contracts at discrete times as needed.
Before disruption, we use $N=100^3,120^3,140^3$ points for the 3 different resolutions. During merger, we use instead $120^3,140^3,160^3$ points. For configurations
in which the black hole spin is aligned with the orbital angular momentum, we only evolve the region above the orbital plane and impose symmetry conditions on that plane
(the number of grid points in the direction orthogonal to the orbital plane is then divided by 2).

Compared to the simulations performed in~\cite{2012PhRvD..85d4015F}, we also fixed an error in the algorithm responsible for communicating source terms
between the two numerical grids, which could cause the time stepping algorithm to be effectively of lower order than expected. The correction leads to reduced errors in the observed trajectories
and in the phase of the gravitational waveforms (see Sec.~\ref{sec:gwacc}). 

\section{Initial Configurations}
\label{sec:id}

The initial data for these simulations is constructed as described in Foucart et al.(2008)~\cite{Foucart:2008a}. The constraints that Einstein's equations impose on the initial variables
are solved in the extended conformal thin sandwich approximation~\cite{York:1999}, under the assumption that the system is in equilibrium in a frame rotating at angular velocity $\Omega_0$
and contracting with radial velocity $-v_r$. The values of $\Omega_0$ and $v_r$ determine the eccentricity of the system, and are chosen iteratively in order to 
minimize that eccentricity~\cite{Pfeiffer:2007a}. We go through two iterations of that procedure, starting from quasi-circular orbits (i.e. initial data with $v_r=0$ and $\Omega_0$ chosen
so that the initial motion of the neutron star center has no radial component). The residual eccentricities at the end of the iterative procedure are $e\sim 0.002-0.004$. 
The free variables in the initial data (conformal
metric, extrinsic curvature) are the weighted superposition of an isolated black hole in Kerr-Schild coordinates and of an isolated neutron star in isotropic coordinates, following the method
developed by Lovelace et al.(2008)~\cite{Lovelace:2008a} for binary black holes. 
A more detailed description of the modifications required to apply this method to black-hole--neutron-star systems is given
in Foucart et al. (2008)~\cite{Foucart:2008a}. 

\begin{table}
\caption{
Initial configurations studied. All binaries have a mass ratio of 1:7, with the black hole dimensionless spin magnitude being 
$\chi_{\rm BH}=0.9$. $\Theta_{\rm BH}$ is the angle between the rotation axis of the black hole and the initial orbital
angular momentum of the binary, $C_{\rm NS}=M_{\rm NS}/R_{\rm NS}$ is the compactness of the neutron star, and 
$R_{\rm NS}^{M=1.4M_\odot}$ the Schwarzschild radius of that neutron star assuming an ADM mass in isolation of $1.4M_\odot$. 
The orbital parameters are the eccentricity $e$, the
initial orbital angular velocity times the total mass of the system, $M\Omega(t=0)$, and the number of gravitational-wave
cycles before the peak of the wave amplitude, $N_{\rm cycles}$ (approximate numbers
given for the precessing systems, in which mode mixing makes this variable
ill-defined).
}
\label{tab:sim}
\begin{tabular}{|c||c|c|c|c|c|c|}
\hline
Name & $\Theta_{\rm BH}$ & $C_{\rm NS}$ & $R_{\rm NS}^{M=1.4M_\odot}$ & $M\Omega_{\rm orbit}^{t=0}$ & $N_{\rm cycles}$ & $e(t=0)$ \\
\hline
R12i0 & $0^\circ$ & 0.170 & 12.2 km & 0.0413 & 20.5 & 0.004 \\ 
R13i0 & $0^\circ$ & 0.156 & 13.3 km & 0.0413 & 20.3 & 0.003 \\
R14i0 & $0^\circ$ & 0.144 & 14.4 km & 0.0413 & 19.7 & 0.003 \\
R14i20& $20^\circ$& 0.144 & 14.4 km & 0.0412 & $\sim 18$ & 0.003 \\
R14i40& $40^\circ$& 0.144 & 14.4 km & 0.0413 & $\sim 17$ & 0.004 \\
R14i60& $60^\circ$& 0.144 & 14.4 km & 0.0415 & $\sim 14$ & 0.002 \\
\hline
\end{tabular}
\end{table}

Two series of initial configurations are considered in this paper (see Table~\ref{tab:sim}), chosen in order to study separately the effects of the radius of the neutron star and of the 
orientation of the black hole spin on BHNS mergers at realistic mass ratios. All configurations consider a black hole of mass $M_{\rm BH} = 7M_{\rm NS} \sim 10M_\odot$
and spin $\chi_{\rm BH}=J_{\rm BH}/M^2_{\rm BH}=0.9$, where $M_{\rm BH}$ is the Christodolou mass of the black hole, $J_{\rm BH}$ its angular momentum, and 
$M_{\rm NS}$ the ADM
mass in isolation of a neutron star with the same baryon mass $M^b_{\rm NS}$ as the neutron star in the binary. The neutron star is initially nonspinning. 
As our initial data does not exactly represent a BHNS binary in quasi-equilibrium, small transients are always observed at the beginning of the simulations. After
those transients, the mass and spin of the black hole are slightly modified (by $\sim 0.01\%$).
All simulations describe the neutron star matter as an ideal fluid with stress-energy tensor $T_{\mu \nu} = (\rho_0 (1+\epsilon) + P) u_{\mu} u_{\nu} + P g_{\mu \nu}$,
and use a $\Gamma$-law equation of state of index $\Gamma=2$:
\beqn
\label{eq:eosP}
P &=& \kappa \rho_0^2 +\rho_0 T\\
\label{eq:eosE}
\epsilon &=& \frac{P}{\rho_0}
\eeqn
where $\rho_0$ is the baryon density of the neutron star material, $P$ its pressure, $\epsilon$ its internal energy, $u_{\mu}$ its 4-velocity, $\kappa$ a free constant and $T$ a variable
related to the temperature of the fluid ($P_{\rm th}=\rho_0 T$ is the thermal pressure in the fluid). To obtain the physical temperature $T_{\rm phys}$ of the fluid
from the variable $T$, we assume that the thermal pressure is the composition of an ideal gas component and a black body component:
\beq
T = \frac{3kT_{\rm phys}}{2m_n} + f\frac{aT_{\rm phys}^4}{\rho_0}
\eeq
where $m_n$ is the nucleon mass, $k$ the Boltzmann constant, and $f$ a function of $T$ reflecting the fraction of relativistic particles in the gas
(see~\cite{Shibata:2007zm,Etienne:2008re} for details).

In the first group of simulations, we consider black hole spins aligned with the orbital angular momentum of the system, and modify the radius of the 
neutron star between $R=12.2\,{\rm km}$ and $R=14.4\,{\rm km}$ (for $M_{\rm NS}=1.4M_\odot$). This is done by modifying the value of the free parameter $\kappa$ in the equation of state
of the fluid. We only consider this simple variation of the equation of state as we know that, to first order, the radius of the neutron star is the most important contribution to the dependence of
BHNS mergers on the equation of state of the fluid~\cite{Duez:2010a}, while the tidal deformability $\lambda\sim k_2 R_{\rm NS}^5$ determines tidal effects during the inspiral~\cite{2010PhRvD..81l3016H}
($k_2$ being the tidal Love number of the neutron star). With respect to these parameters, the configurations considered here are within the range currently allowed for real neutron stars.
They do however fail to reproduce other properties of neutron stars which are not as relevant to this study. For example, all three equations of state have a maximum mass 
smaller than $2M_\odot$ (which is more important for studies of stellar collapse and of the evolution of hypermassive neutron stars than for tidal disruption and BHNS mergers, in which the maximum density of the fluid only decreases over the course of the evolution), 
they have a very simplified temperature dependence, and they do not describe the composition of the fluid.  
The largest neutron star (case R14i0) is identical to the configuration studied in Foucart et al.(2012)~\cite{2012PhRvD..85d4015F}, except
that the initial separation is larger than in our previous work. As we will see, these parameters study most of the transition between BHNS mergers resulting in the formation of 
massive disks and those having nearly no material left out of the black hole a few milliseconds after merger. 

As we are considering $\Gamma$-law equations of state, it is worth mentioning that any of these simulations actually represent a continuum of systems, as they are invariant through the rescaling
\beqn
M' &=& K*M\\
R' &=& K*R\\
T' &=& K*T
\eeqn
where $M$ is the mass scale of the binary, $R$ the distance scale, $T$ the time scale, and $K$ an arbitrary positive constant. 
A more useful description of the initial conditions would thus use quantities which are also
invariant under this rescaling, i.e. the mass ratio $q=M_{\rm NS}/M_{\rm BH}$, the compactness of the neutron star $C_{\rm NS}=M_{\rm NS}/R_{\rm NS}$, and the dimensionless time $\tau=T/M$
(in units in which $G=c=1$).
Realistic neutron stars of $1.4M_\odot$ probably have radii in the range $R_{\rm NS} \sim 9\,{\rm km}-14\,{\rm km}$~\cite{2010PhRvL.105p1102H}, 
with the most likely values being $R_{\rm NS}\sim 11\,{\rm km}-12\,{\rm km}$~\cite{2010ApJ...722...33S}. 
The values of $C_{\rm NS}$ considered here are thus more likely to be found in neutron stars of ADM mass around or slightly below $1.4M_\odot$, 
while probably unrealistically low for very massive neutron stars ($M_{\rm NS} \sim 2M_\odot$).

The second set of simulations considers variations of the orientation of the black hole spin while maintaining the equation of state fixed (using the larger neutron star with $C_{\rm NS}=0.144$).
In terms of disk formation, this choice of equation of state is clearly optimistic, although not unrealistic, and thus provides an upper bound on the mass remaining outside of the black
hole after merger. Moderate misalignments of the spin of the black hole with respect to the angular momentum of the binaries are an expected consequence of the kick that the supernova explosion
is likely to impart to the forming neutron star. The actual distribution of the misalignment angle $\Theta_{\rm BH}$ between the spin of the black hole and the orbital angular momentum of the
binary is currently unknown, although $\Theta_{\rm BH}\lo 90^\circ$ should probably be favored~\cite{2008ApJ...682..474B}. 
Here we vary $\Theta_{\rm BH}$ between $0^\circ$ and $60^\circ$, as such misalignments
are both physically realistic and covering the range of parameters over which the properties of the final remnant vary significantly (at least for BHNS systems of mass ratio $q=7$ with black hole
spins $\chi_{\rm BH}=0.9$). Misalignments are also often quoted as the angle $\eta_{\rm BH}$ between the black hole spin and the total angular momentum of the system. For the systems
considered here, $\Theta_{\rm BH}=(20^\circ,40^\circ,60^\circ)$, we have $\eta_{\rm BH}=(7^\circ,14^\circ,21^\circ)$.

All simulations begin at a coordinate separation $d=7.44M$, where $M=M_{\rm BH}+M_{\rm NS}$ is the total ADM mass of the system at infinite separation. 
This corresponds to an initial orbital angular velocity 
$M\Omega_{\rm orbit}(t=0)\sim 0.041$, or an initial gravitational-wave frequency $f_{\rm GW}(t=0)\sim 235\,{\rm Hz} \left(\frac{1.4M_\odot}{M_{\rm NS}}\right)$. Over the course of the 
simulation, the binary will go through $7-10$ orbits before merging.

\section{Accuracy}
\label{sec:acc}

The combination of spectral and finite difference methods used in our simulations can make it difficult to obtain strict
error estimates: spectral methods are exponentially convergent in regions in which all variables are smooth, but only show
polynomial convergence in the presence of discontinuities (such as at the surface of the neutron star or at a shock front). 
The finite difference methods used to evolve the equations of relativistic hydrodynamics are second order in smooth regions, and 
first order at the location of shocks. As the region in which we get first order convergence should be of measure zero, we 
expect at least second-order convergence as we increase the resolution of the finite difference grid. 
In practice, we generally observe much faster convergence between the 3 resolutions considered here, particularly for 
quantities evolved on the spectral grid (e.g. trajectories, gravitational-wave signal,...). A conservative estimate of our error 
would thus be to assume second order convergence between our medium and high resolutions - the actual error being somewhere
between that value and the optimistic estimate obtained by simply looking at the difference between those two simulations. 
The ratio between these pessimistic and optimistic error estimates is $\sim 3$ for the simulations presented here.

\subsection{Final Remnant}

Of the characteristics of the final remnant listed in Table~\ref{tab:rem}, the parameters of the black hole (mass and spin) are 
the most accurate, with relative errors $\epsilon^{\rm rel}_{\rm BH} \lo 0.3\%$ (i.e. differences of $\sim 0.1\%$ between the medium
and high resolutions). Global mass measurements (disk mass, tidal tail mass) are already less accurate,
with $\epsilon^{\rm rel}_{\rm Mass} \lo 15\%$ ($0.01M_{\rm NS}$ difference measured in the final remnant mass of the medium and 
high resolution runs for configuration R13i0). Finally, the maximum
density within the disk is only order-of-magnitude accurate: the distribution of matter within the disk remains fairly asymmetric
and time-dependent at the end of the simulation, and variations of $\sim 50\%$ within $\sim 1\,{\rm ms}$ should still be expected.
Measurements of the mass of unbound material and the properties of this ejecta have similar errors, and are discussed in
more details in Sec.~\ref{sec:ej}.

\subsection{Waveform Accuracy}
\label{sec:gwacc}

\begin{figure}
\includegraphics[width=8.2cm]{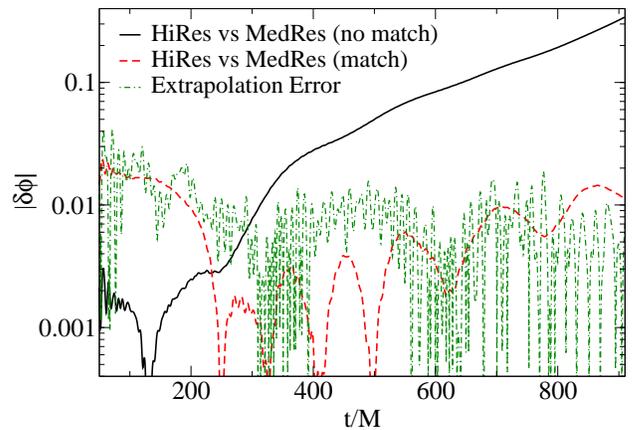}
\caption{Phase error in the dominant (2,2) mode of the gravitational-wave signal for simulation R13i0. We show the phase difference between our
standard and high resolutions, both with (dashed red line) and without (solid black line) aligning the waves in phase and time. The dash-dotted green
curve shows an estimate of the error in the extrapolation method (obtained by comparing extrapolation using different polynomial orders).}
\label{fig:gwerr}
\end{figure}

The phase accuracy of the gravitational waveforms in the non-precessing simulations presented here is about a factor of 2 better
than in our last set of simulations~\cite{2012PhRvD..85d4015F}, even though the evolutions are $\sim 2-3$ orbits longer. This is most likely
due to the correction of an error which effectively decreased the order of the time stepping method used in our simulations. 
Fig.~\ref{fig:gwerr} shows the phase difference between the medium and high resolution of the inspiral of simulation R13i0, 
both without any matching (i.e. by directly computing the phase difference between the output of the 2 resolutions), and after 
matching the waveforms over one period of the radial oscillation of the orbit, choosing a time and phase shifts minimizing the
difference between the two waveforms in the matching region. The first method is the most direct assessment of the effect of
numerical errors on the phase of the gravitational-wave signal, which are here of the order of a few tenths of a radian
during the inspiral. 
The second method is more useful when comparing waveforms obtained in simulations starting from different initial conditions,
and shows how different the waveforms would look to a gravitational-wave detector. Fig.~\ref{fig:gwerr} shows that, for matched
waveforms, differences of the order of a few percents of a radian or less cannot be resolved by our numerical simulations.

When considering waveform accuracy, numerical errors due to the discretization of the evolution equations are however 
only part of the problem. Another potential source of error comes from extracting gravitational waves at finite radii, 
and then using 
polynomial extrapolation to obtain the waveform at null infinity~\cite{2009PhRvD..80l4045B}. An estimate of the error due to this process 
can be obtained by comparing the waveforms obtained using different polynomial orders for the extrapolation. Fig.~\ref{fig:gwerr}
shows that this error is $\sim 0.01\,{\rm rad}$. Phase differences of the same order can also be due to the eccentricity of the
binaries, at least for the eccentricities $e\sim 0.002-0.004$ considered here. This can easily be seen from
the oscillations in the phase difference between different configurations shown in Fig.~\ref{fig:gwphasenp} (the oscillations
in the phase difference between simulations R13i0 and R14i0 are smaller that those between R12i0 and R14i0 because the radial
oscillations of the first two cases happen to be nearly in phase at the beginning of the simulation).

Adding all these sources of error, we can thus estimate that differences between numerical waveforms are large enough to
be measured at our current accuracy only if, for matched waveforms, we have $\delta \phi \ge 0.05\,{\rm rad}$ during inspiral.

\section{Results}
\label{sec:nr}

\subsection{Non-Precessing Binaries}

\subsubsection{Inspiral : Tidal Effects}
\label{sec:tides}

Before the disruption of the neutron star, the main differences between a BHNS inspiral and a BBH inspiral are due to the finite
size of the neutron star, and its distortion under the influence of the tidal field of the black hole. The tidal distortion
of the neutron star, and in particular its effect on the gravitational-wave signal, has already been studied in the 
Post-Newtonian(PN) framework. During the early inspiral, Hinderer et al.~\cite{2010PhRvD..81l3016H} found that for BNS systems the tidal
effects would only be detectable by Advanced LIGO for the most favorable configurations (i.e. the largest neutron stars, 
see also~\cite{PhysRevD.84.104017} for similar results considering the tidal effects up to the disruption of the neutron star). 
Over the last few
orbits, Damour et al.~\cite{2012PhRvD..85l3007D} find that tidal parameters would be detectable for BNS mergers 
of moderate signal-to-noise ratio ($\rho \sim 16$). 
But as these effects are significantly smaller for more asymmetric mass ratios, the detection of tidal effects 
through gravitational waves is much more difficult for BHNS 
systems. A more detailed discussion of the detectability of the neutron star equation of state in our mergers is
offered in Sec.~\ref{sec:npgw} - but from Fig.~\ref{fig:gwphasenp} alone, where we show the phase difference between our 3
non-precessing simulations, it is easy to see that up to 4 gravitational-wave cycles before the peak of the 
gravitational-wave signal ($f_{\rm GW}\lo 500\,{\rm Hz}$) 
the difference between these cases is not resolved numerically. 
This could however be due either to the fact that tidal effects on the waveform are extremely small,
or to a failure of the simulations to capture the tidal distortion of the neutron star properly. Accordingly, we need to test
that the neutron star is tidally distorted during inspiral, and that these tidal effects scale as expected. We compute the quadrupole moments
of the neutron star
\beq
Q_{ij} = \int \rho \left(x_i x_j - \frac{1}{3} \delta_{ij} r^2\right) dV
\eeq
(where $\rho=\sqrt{g}W\rho_0$, $W=\sqrt{1+g^{ij}u_iu_j}$, $u_i$ is the spatial component of the 4-velocity and $dV$ is the
volume element, so that $\int \rho dV = M^b_{\rm NS}$), 
and assume that they are due to first order to the composition of the tidal distortion of the neutron star and of a coordinate boost, acting along
orthogonal directions. $Q_{ij}$ is then diagonal in the coordinate frame $(\hat{e}_\pm, \hat{e}_z)$, where $\hat{e}_\pm$ are two 
orthogonal unit vectors in the equatorial plane of the binary and $\hat{e}_z$ is a unit vector in the direction of the orbital 
angular
momentum (by symmetry, $Q_{xz}=Q_{yz}=0$). The orientation of $\hat{e}_{\pm}$ in the equatorial plane is a priori unknown, and
practically determined by solving for the rotation matrix diagonalizing $Q_{ij}$.
To first order, we assume that the tidal distortion causes the neutron star to stretch along the 
direction $\hat{e}_+$ and contract along $\hat{e}_-$ and $\hat{e}_z$, while the boost causes a contraction along $\hat{e}_-$ and a 
stretch along $\hat{e}_+$ and $\hat{e}_z$, i.e.
\beqn
Q_{++} &=& \frac{2Q}{3} + \frac{B}{3} \\
Q_{--} &=& - \frac{Q}{3} - \frac{2B}{3} \\
Q_{zz} &=& -\frac{Q}{3} + \frac{B}{3}
\eeqn 
where $Q$ is the magnitude of the tidal distortion, and $B$ the amplitude of the boost distortion (by
construction, $Q_{++}+Q_{--}+Q_{zz}=0$. So we can solve any 2 of the 3 equations above for the unknowns $B$
and $Q$, and the third will automatically be satisfied). From this decomposition, we can
also retrieve the lag angle $\Psi_{\rm tidal}$ between the tidal bulge and the line connecting the center of the hole and the center of
the neutron star (i.e. the angle between $\hat{e}_+$ and the line connecting the two centers).

\begin{figure}
\includegraphics[width=8.2cm]{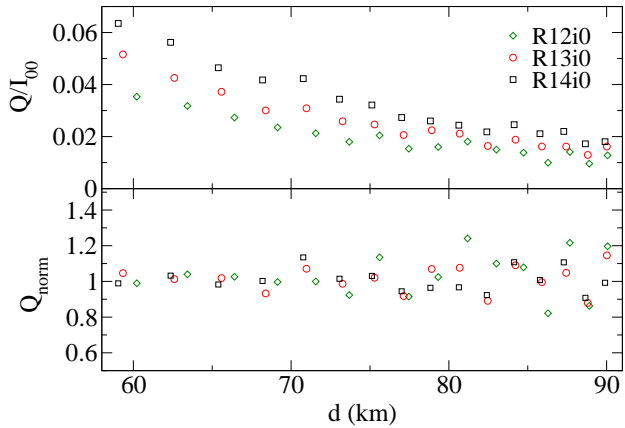}
\caption{{\it Top:} Tidal quadrupole Q normalized by $I_{00}$. {\it Bottom:} Normalized quadrupole $Q_{\rm norm}$ (see eq.~\ref{eq:qnorm}).}
\label{fig:quad}
\end{figure}

These quantities are clearly dependent on the coordinate system chosen. We cannot entirely remove that dependence, but can at
least get a reasonable normalization for the quadrupole moments from the quantity
\beq
I_{00}=\int \rho r^2 dV.
\eeq
For all simulations, we find similar boost components $B/I_{00} \sim 2\%$: $B/I_{00}$ is a function of the binary separation, but does not depend on the equation of state considered 
(for a true Lorentz boost of a spherical star, we should get $B/I_{00}\sim (v/c)^2$). 
The lag angle is fairly constant too, with $\Psi_{\rm tidal}\sim 20^\circ-25^\circ$
for separations $d \sim 60\,{\rm km}-90\,{\rm km}\sim 4M_{\rm BH}-6M_{\rm BH}$. The tidal component, on the other hand, varies strongly with both the binary separation and the equation
of state. In Fig.~\ref{fig:quad}, we show $Q/I_{00}$ for the three different equations of state considered here, and in the 
range of binary separations $d \sim 60\,{\rm km}-90\,{\rm km}$. The tidal distortion goes from being of the same order as the boost
effect at $d\sim90\,{\rm km}$ to a factor of $2-3$ larger at $d\sim 60\,{\rm km}$. Not surprisingly, the larger neutron star is significantly more distorted.

To leading order, we expect the tidal distortion $Q$ and the tidal field of the black hole 
($\sim M_{\rm BH}/d^3$) to be related by
\beq
Q \sim 2 k_2 R_{\rm NS}^5 \frac{M_{\rm BH}}{d^3}
\eeq
where $k_2$ is the tidal Love number of the neutron star
(for $\Gamma=2$ polytropes, which at $T=0$ are equivalent to the $\Gamma$-law equation of state used in our simulations,
$k_2$ was computed by Hinderer~\cite{2008ApJ...677.1216H}).
To verify that the tidal effects scale as expected, we thus compute the normalized tidal parameter $Q_{\rm norm}$:
\beq
\label{eq:qnorm}
Q_{\rm norm} = \frac{Q}{0.0071 I_{00}^{R14}} \frac{k_2^{R14}}{k_2} \left(\frac{C_{\rm NS}}{0.144}\right)^5 \left(\frac{d}{d_0}\right)^3.
\eeq
The superscript $R14$ refers to values for simulation R14i0, and $d_0 = 123\,{\rm km}$ is the initial separation of the binary.
The normalization $I_{00}^{R14}$ is the value of $I_{00}$ for simulation R14i0 at the separation $d$ at which we are
measuring $Q_{\rm norm}$. The numerical factor $0.0071$ is computed in the limit 
$d \rightarrow \infty$ (i.e. with $I_{00}^{R14}$ computed for an isolated neutron star), so that $Q_{\rm norm}(d\rightarrow \infty) =1$.
The scalings of $k_2$, $C_{\rm NS}$ and $d$ are chosen so that, as long as the tidal distortion of the neutron star follows the theoretical
predictions, we will measure $Q_{\rm norm}=1$. 

In practice, our ability to measure $Q_{\rm norm}$ accurately is limited by the fact that the boost $B$ and normalization $I_{00}$ 
only approximately model the distortion of the NS due to coordinate effects (i.e. the boost, but also any other gauge effect due
to the coordinate choices made in the simulation). 
Measurements of $Q_{\rm norm}$ are thus unreliable 
for $Q\lo B \sim 0.02$. Figure~\ref{fig:quad} shows that for $Q\sim B$, the scatter in the measurement of $Q_{\rm norm}$
is $\sim 30\%$, while for $Q\sim 2B-3B$, it decreases below $10\%$. All measurements of $Q_{\rm norm}$ are compatible
with $Q_{\rm norm}=1$ within that scatter, thus showing that the tidal distortion of the neutron star follows approximately
the predictions of Ref.~\cite{2008ApJ...677.1216H}, even at close separations.

From these computations, we can thus conclude that the tidal distortion of the neutron star during the late inspiral is
resolved by our numerical simulations, and scales with the binary separation and the equation of state of the neutron star
in the manner expected from theoretical calculations, at least within $\delta Q/I_{00}\sim 0.005$.

\subsubsection{Merger and Disk Formation}

\begin{table}
\caption{
Properties of the final remnant. $M_{\rm remnant}^{5ms}$ is the baryon mass remaining outside of the black hole 5ms after merger. $M_{\rm tail}^{5ms}$ is the baryon mass
located at a coordinate radius greater than $\sim 200\,{\rm km}$ at the same time. $\rho_{\rm max}^{5ms}$ is the maximum density in the disk, $\chi^f_{\rm BH}$
the dimensionless spin of the black hole at the end of the simulation, $M_{\rm BH}^{\rm f}$ the final Christodolou mass
of the black hole and $M$ the ADM mass of the system at infinite separation.
}
\label{tab:rem}
\begin{tabular}{|c||c|c|c|c|c|}
\hline
Name & $\frac{M_{\rm remnant}^{5ms}}{M^b_{\rm NS}}$ &  
$\frac{M_{\rm tail}^{5ms}}{M^b_{\rm NS}}$ & $\rho_{\rm max}^{5ms}$ [$10^{11}\,{\rm g/cm}^3$] 
& $\chi^f_{\rm BH}$ & $\frac{M_{\rm BH}^{\rm f}}{M}$ \\ 
\hline
R12i0 & 0.10 & 0.06 & 2 & 0.923 & 0.960 \\ 
R13i0 & 0.20 & 0.11 & 3 & 0.919 & 0.950\\
R14i0 & 0.30 & 0.16 & 21 & 0.910 & 0.935 \\
R14i20& 0.28 & 0.15 & 17 & 0.909 & 0.939 \\
R14i40& 0.15 & 0.10 & 3 & 0.898 & 0.959\\
R14i60& 0.03 & 0.03 & 0.4 & 0.862 & 0.978\\
\hline
\end{tabular}
\end{table}

An important question when considering BHNS mergers is the form of the post-merger remnant. To first order, this depends
on whether the neutron star is disrupted before reaching the innermost stable circular orbit of the black hole or not. In
the first case, a large amount of matter can remain outside of the hole after merger in the form of an accretion disk
and a tidal tail. In the second case, no matter will remain. For a BHNS merger to be the progenitor of a short gamma-ray
burst, the creation of an accretion disk is necessary. Accordingly, SGRBs are only possible if the neutron star disrupts.  
Stellar disruption is facilitated by low black hole masses, high black hole spins and large neutron stars 
(see~\cite{2012arXiv1207.6304F} for a simple fit to the results of previous numerical simulations). 
At low mass ratios ($M_{\rm BH}/M_{\rm NS}<5$), 
a moderately spinning black hole $\chi_{\rm BH}\sim 0.5$ is generally sufficient to provide disks of $\sim 0.1M_\odot$.
For the more massive black holes considered here, however, this is no longer the case. We have already shown that for spins
$\chi_{\rm BH} \leq 0.7$ disk formation is unlikely even for large neutron stars ($R_{\rm NS}=14.4\,{\rm km}$). The simulations presented
here begin to explore how smaller neutron stars fare. 

\begin{figure*}
\includegraphics[width=6cm]{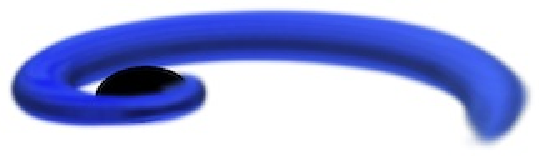}
\includegraphics[width=6cm]{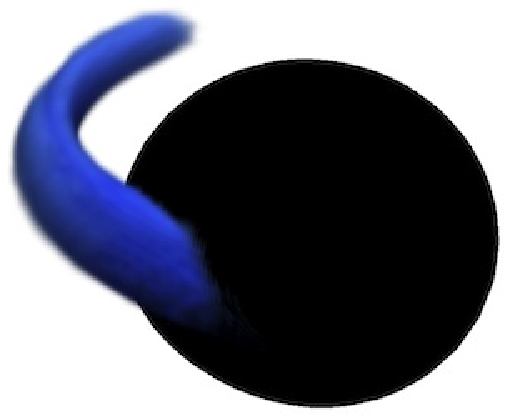}\\
\includegraphics[width=6cm]{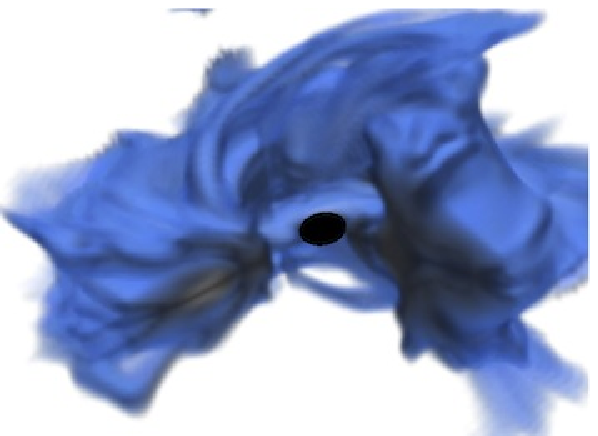}
\includegraphics[width=6cm]{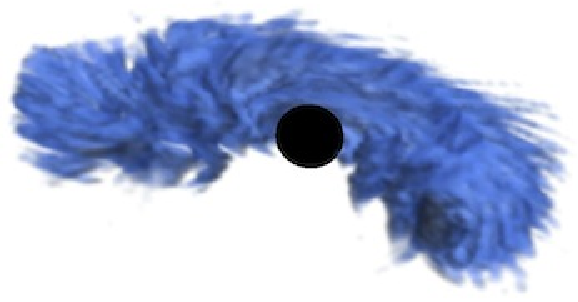}
\caption{Merger for the non precessing cases R14i0 (left) and R12i0 (right). The top panel shows the system at the time
at which half of the neutron star material has been accreted onto the black hole. We show densities down to 
$\rho_{\rm min}\sim 10^{-7}M_\odot^{-2} \sim 6 \times 10^{10}\,{\rm g/cm^3}$. The bottom panel shows the remnant 5ms later, 
plotting densities down to $\rho_{\rm min}\sim 10^{-8}M_\odot^{-2} \sim 6 \times 10^{9}\,{\rm g/cm^3}$ and cutting out
the $x>0$, $y<0$ quadrant. The differences in scale between the 4 figures can be determined knowing that the 
size of the black hole is always $R_{\rm BH}\sim 15\,{\rm km}$.}
\label{fig:simnp}
\end{figure*}

Fig.~\ref{fig:simnp} shows snapshot of simulations R14i0 and R12i0, the largest and smallest neutron stars considered here, both in the
middle of the stellar disruption, and $5\,{\rm ms}$ later. The larger neutron star disrupts far enough from the black hole for a large
portion of the matter to be initially ejected into a tidal tail, but the smaller neutron star disrupts just outside of the ISCO of the black
hole. The top-right panel of Fig.~\ref{fig:simnp} in particular shows how close the smaller neutron star is to the hole when it
disrupts. From this picture, the fact than any material remains outside of the hole after merger is surprising in itself, and
an indication of how strong the effects of the black hole spin can be on infalling material. 

\begin{figure}
\includegraphics[width=8.2cm]{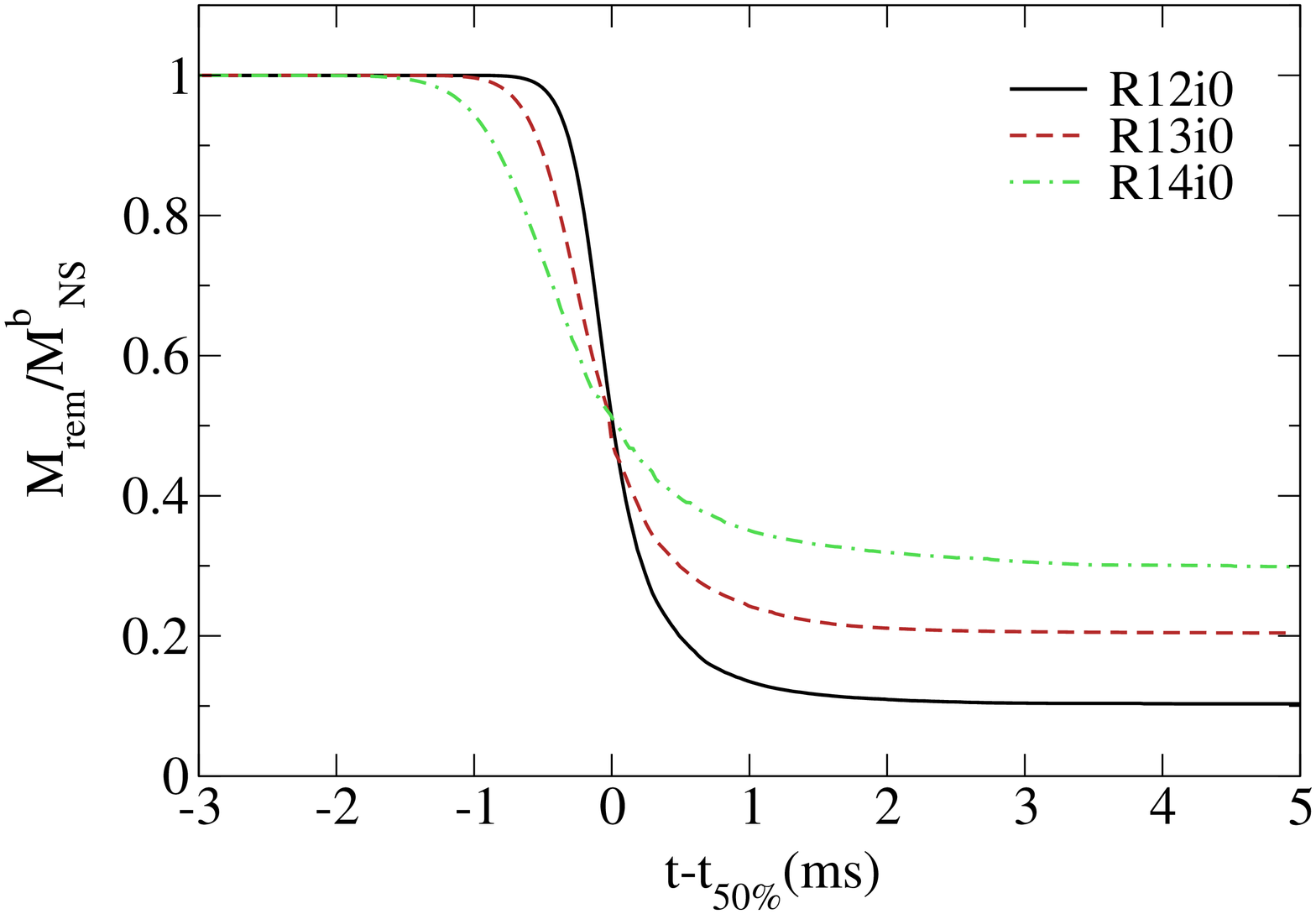}
\caption{Baryon mass remaining outside of the black hole as a function of time, for the 3 non-precessing cases R14i0, R13i0 and R12i0.
We shift all the curves by the time $t_{50\%}$ at which half of the matter has been accreted onto the black hole.}
\label{fig:massnp}
\end{figure}

\begin{figure}
\includegraphics[width=8.2cm]{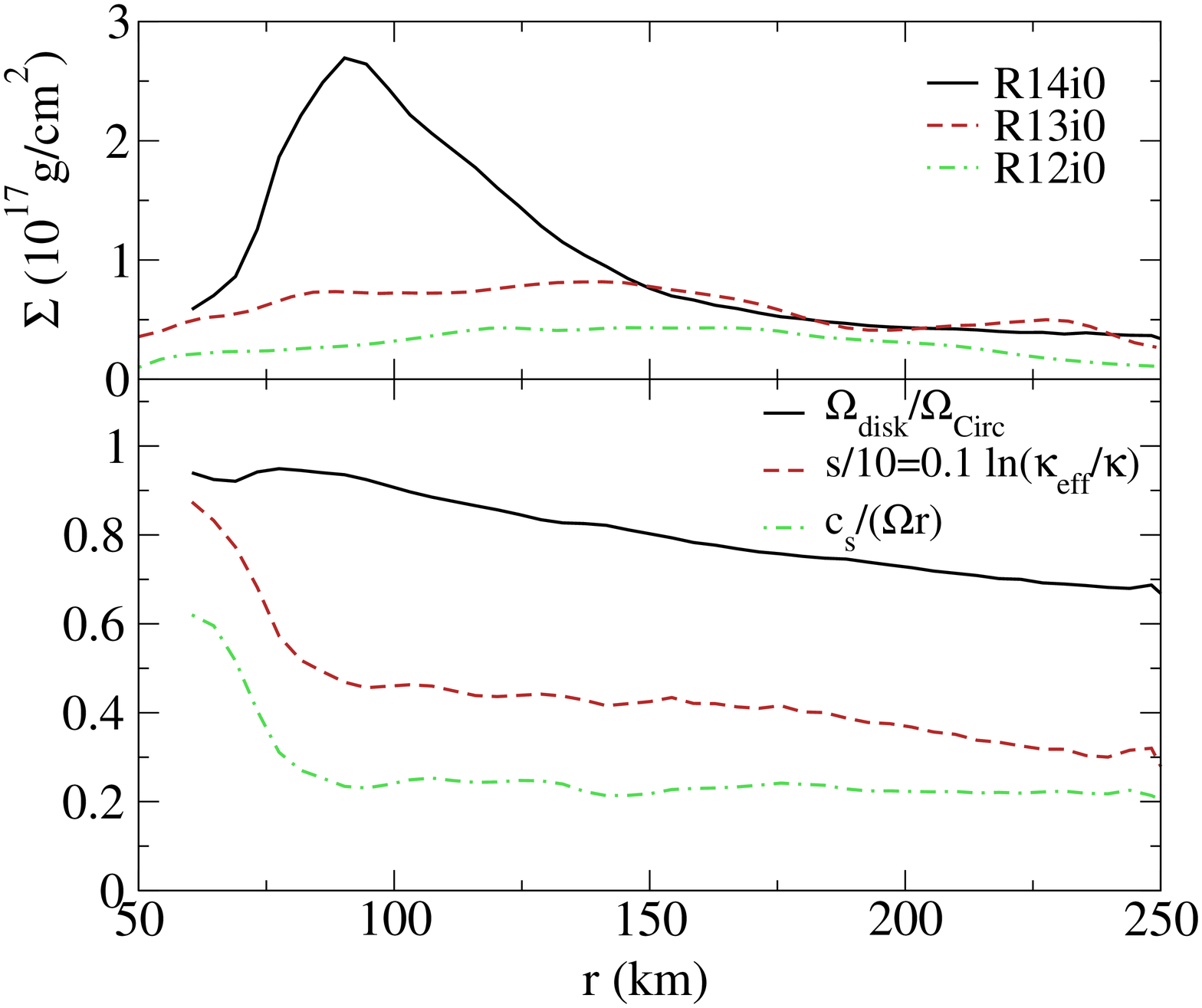}
\caption{{\it Top:} Average surface density 5ms after merger for the 3 non-precessing cases R14i0, R13i0 and R12i0.
{\it Bottom:} Disk profile for simulation R14i0. Shown are the angular velocity (normalized by the velocity for circular
orbits in this metric), the entropy and the sound speed (normalized by $\Omega r$) as a function of the 
circumferential radius $r$.}
\label{fig:diskprof}
\end{figure}

Important differences are observed between the final state of these mergers. For the larger neutron star, 
$30\%$ of the matter remains outside of the black hole $5\,{\rm ms}$ after merger. 
More importantly, about half that material has already formed a thick
accretion disk, of about $100\,{\rm km}$ in radius and with peak density $\rho_{\rm max}^{5ms}\sim 2\times 10^{12}\,{\rm g/cm}^3$.
The formation of a hot, thick disk is less obvious for the smaller neutron stars. The amount of material remaining outside of the 
black hole is by no means negligible ($10\%-20\%$ of the neutron star, see Fig.~\ref{fig:massnp}), but the maximum density 
is about an order of magnitude lower. In fact, if we look at the average surface density as a function of radius 
(Fig.~\ref{fig:diskprof}), we see no evidence of an accumulation of higher density material at lower radii ($\sim 100\,{\rm km}$),
while that feature is clearly visible for the larger neutron star. 
From these results, we can also infer that smaller neutron stars
$R_{\rm NS}\lesssim 11\,{\rm km}$ would probably be unable to form any long-lived remnant.

Evolutions including all the necessary microphysics (magnetic fields, neutrino cooling) will be necessary to determine
how these disks evolve over longer time scales ($\go 0.1s$). 
We can already see, however, that for all three configurations the material remaining outside of the black hole is hot ($<T>\sim 2\,{\rm MeV}$),
and would be cooled by neutrino emission. In that respect, the differences in density between the remnants could be significant, as they
modify the opacity of the disk to neutrino radiation, and thus the efficiency with which the disk can transfer its energy into 
neutrinos.

The properties of the final remnant are presented in Table~\ref{tab:rem}. Comparing our results for the amount of material
remaining outside of the black hole at late times with the predictions of Ref.~\cite{2012arXiv1207.6304F}, 
we find good agreement (within $2\%-3\%$ of the neutron star mass) for the two smallest stars. The largest star
forms a disk heavy enough that we are out of the range in which the predictions of Ref.~\cite{2012arXiv1207.6304F}
are expected to be valid --- and indeed, the disk formed in the simulation is significantly more massive that what 
Ref.~\cite{2012arXiv1207.6304F} would predict. We also find consistency between our simulations
and Ref.~\cite{2012arXiv1207.6304F} on the neutron star radius below which no matter will remain outside of the
black hole after merger ($\sim 10.5\,{\rm km}-11\,{\rm km}$). A more careful examination of the differences between our numerical
results and Ref.~\cite{2012arXiv1207.6304F} indicates that in the regime of high spin, high black hole masses considered here, 
the remnant mass probably has a steeper dependence on the radius of the neutron star than what would be guessed from Ref.~\cite{2012arXiv1207.6304F}, 
a model fitted mostly to lower mass systems. However,
these differences could also be explained by the expected variations in the remnant mass due to the internal structure of the neutron star (i.e. the fact
that two neutron stars with the same radius but different internal structure will result in different post-merger disk masses), especially
considering the fact that all of the simulations used to fit the model in~\cite{2012arXiv1207.6304F} had larger tidal Love number $k_2$ than the
neutron stars from simulations R13i0 and R12i0. Overall, the magnitude of the differences between the numerical results and the model
are roughly at the expected level. The final black hole masses and spins are also within
the expected errors of existing analytical models, i.e. $1\%-2\%$ away from the values derived by Pannarale~\cite{2012arXiv1208.5869P}. 

For all configurations, the disruption and merger of the neutron star occurs over $\sim 2\,{\rm ms}$ (see Fig.~\ref{fig:massnp}). Mass accretion at
later times is negligible compared with what is observed in lower mass ratio systems. When a disk forms, its main characteristics are
however fairly similar to the lower mass ratio cases --- except of course for the aforementioned lower densities and larger disk radii,
which are a natural consequence of the higher black hole mass. Fig.~\ref{fig:diskprof} shows a few characteristics
of the forming accretion disk for the most strongly disrupted case. The surface density peaks at a distance of $100\,{\rm km}$ from the
black hole, and the disk extends to about $150\,{\rm km}$
\footnote
{
  Distances are measured in terms of the circumferential radius in the equatorial plane,
  $r_{\rm circ}=\frac{1}{2\pi}\int_0^{2\pi}\sqrt{g_{\phi\phi}}d\phi$,
  where $\phi$ is the azimuthal angle.
}. 
The orbital velocity profile is slightly sub-Keplerian (by about $10\%$), while
the sound speed is $\sim 0.25 \Omega r$ and thus compatible with a thick, thermally supported disk of scale height 
$H=0.25r$. This is consistent with the actual scale height of the disk, $H \sim 0.2r-0.3r$. Finally, the inner edge of the disk is 
particularly hot: we plot an estimate of the entropy
\beq
s = \ln\left(\frac{\kappa_{\rm eff}}{\kappa}\right) = \ln \left(\frac{P}{P(\rho_0,T=0)}\right)
\eeq
(the effective constant $\kappa_{\rm eff}$ is defined by $P = \kappa_{\rm eff} \rho_0^2$), and find $s\sim 9$ for 
$r\sim 60\,{\rm km}$. Within the disk, we still have $s\sim 4-5$. As the disk settles down over $\sim 10\,{\rm ms}-20\,{\rm ms}$, we would
expect the entropy to exhibit a minimum at the peak of the surface density distribution, as was observed in lower mass ratio 
systems~\cite{Duez:2010a,2011PhRvD..83b4005F}.
 
 \subsubsection{Ejecta}
\label{sec:ej}

The ejection of unbound material by compact binary mergers is a prerequisite for some electromagnetic counterparts, most notably emissions
due to the radioactive decay of the neutron-rich ejecta~\cite{2011ApJ...736L..21R,2012ApJ...746...48M}. This ejecta can be obtained through
various physical processes: unbound material in the tidal tail, but also magnetically-driven~\cite{2011ApJ...734L..36S} or 
neutrino-driven~\cite{2009ApJ...690.1681D} winds. The study of winds goes beyond
the scope of this article, as this requires accurate long-term evolution of the remnant disk and the inclusion of physical processes that are
neglected in this work (magnetic fields, neutrino radiation). We will thus limit ourselves to the measurement of unbound material in the tidal tail.

Even the presence of ejected material in the tidal tail
can be difficult to assess in general relativistic simulations, particularly at high mass ratios. 
Maintaining a high enough resolution in both the disk-forming region and the tidal tail is challenging,
and in practice matter can only be reliably evolved up to a distance of a few hundred kilometers from the black hole. This is indeed one of the main disadvantage of
any grid-based simulations when compared with smoothed particle hydrodynamics methods, which can easily follow the evolution of tidal tails.
In a time-independent spacetime and when pressure forces are negligible, it is easy to determine whether material is unbound: 
if $u_t<-1$, then the material will escape to infinity (and $-u_t$
is the Lorentz factor of the fluid at infinity). This condition is also a fairly good approximation for low-density material far away from the central black hole after
a compact binary merger, but becomes more and more inaccurate as one gets close to the black hole, or densities in the tidal tail become higher.

One way to assess whether using the $u_t<-1$ condition to find unbound material is accurate is to follow material over a sufficiently long period of time,
and check that $u_t$ doesn't vary much. In our simulations, however, this only occurs for $\sim 1\,{\rm ms}$ before the material leaves the numerical grid,
which leads to large uncertainties in the amount of unbound material, and its characteristics. A more detailed discussion of these issues will be presented
in Deaton et al. (in preparation). Here, we limit ourselves to a discussion of measurements of $u_t$ at relative low radii ($\lo 20M_{\rm BH} \approx 300\,{\rm km}$) 
and over short timescales, and note the uncertainties due to these approximations. 

\begin{figure}
\includegraphics[width=8.2cm]{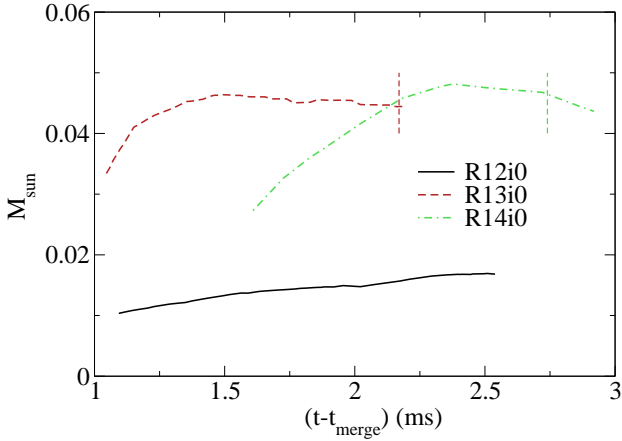}
\caption{Mass of the unbound material, as measured by the condition $u_t<-1$. For each configuration, $t_{\rm merge}$ is the time at which $50\%$ of
the neutron star material has been accreted onto the black hole. The dashed vertical line represent the time at which $0.001M_\odot$ has escaped
the grid (the low density tidal tail of simulation R12i0 cannot be followed
accurately for more than $2.5\,{\rm ms}$, at which point we stop measuring
the mass of the ejecta).}
\label{fig:ej}
\end{figure}

On Fig.~\ref{fig:ej}, we plot the amount of mass with $u_t<-1$ on the grid (and more than $60\,{\rm km}$ away from the black hole). 
The most compact neutron star, simulation R12i0, naturally has the least material in a tidal tail: about $0.09M_\odot$, of which $\sim 0.015M_\odot$ appear
unbound $2.5\,{\rm ms}-3.5\,{\rm ms}$ after accretion onto the black hole begins ($1.5\,{\rm ms}-2.5\,{\rm ms}$ after the time at which $50\%$ of the neutron star material has been accreted onto the black hole),
with variations of only $0.002M_\odot$ over the $1\,{\rm ms}$ period over which measurements appear reliable.
Moving up in stellar radius, simulation R13i0 offers the most reliable measurement of the ejected mass, with a stable value of $M_{ej}^{R13}\sim 0.046M_\odot$ long before
matter starts flowing out of the grid (the total mass of the tidal tail, in this case, is $\sim 0.16M_\odot$), and variations of $0.002M_\odot$ over $1\,{\rm ms}$. 
From this run, we can also estimate the relative error due to finite numerical resolution in these measurements, and get $\epsilon_{ej}<40\%$ (at our ``medium'' resolution, 
we found $M_{ej}^{R13} \sim 0.040M_\odot$). This is distinct from the uncertainties due to the approximation made when using $u_t$ as a proxy for finding whether material is bound or not,
and appears to be the dominant source of error for simulations R12i0 and R13i0.  
Finally, the largest neutron star shows the most uncertain measurements. Velocities and densities in the ejecta are generally higher: the approximate
method takes more time to become accurate, but material remains on the grid for a shorter amount of time. There also is material with $u_t<-1$ flowing directly
out of the forming accretion disk, presumably as a result of shocks during disk formation, which makes it impossible to have all of the potential ejecta in the range
$60\,{\rm km}<r<300\,{\rm km}$ at any given time. Even by expanding the outer boundary of the grid by $50\%$ compared
with the two other runs, we thus find that the measured $M_{ej}^{R14}\sim 0.050M_\odot$ only appear to settle at the time at which matter starts flowing out of the grid 
(and boundary effects might influence the properties of the ejecta). It is thus quite likely that the ejected mass is slightly larger than what is observed in the 
simulation.
Overall, adding the two main sources of error (numerical resolution and use of $u_t$), we estimate that the ejected masses are
\beqn
M_{ej}^{R12} &=& 0.015 M_\odot \pm 0.010 M_\odot \\
M_{ej}^{R13} &=& 0.046 M_\odot \pm 0.020 M_\odot \\
M_{ej}^{R14} &=& 0.050 M_\odot \pm 0.035 M_\odot.
\eeqn

\begin{figure}
\includegraphics[width=8.2cm]{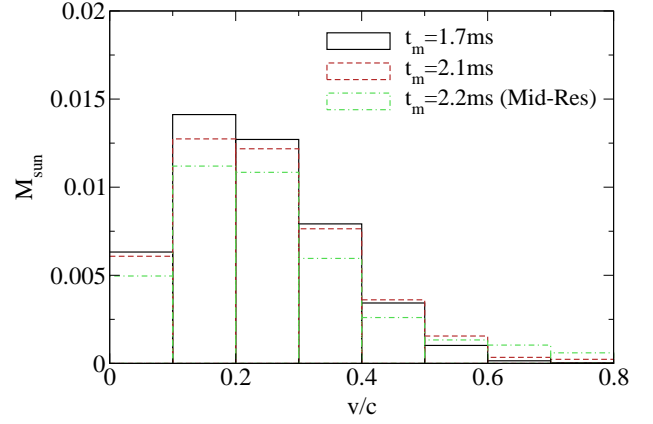}
\caption{Distribution of asymptotic velocities for the unbound material of simulation R13i0. Two different times of the high-resolution simulation, and one time of the medium
resolution simulation are shown.}
\label{fig:velej}
\end{figure}

From the measurements of $u_t$, we can also determine the distribution of the velocity of the fluid at large distance from the black hole. 
This is shown in Fig.~\ref{fig:velej} for configuration R13i0.
The distribution peaks at $v/c\sim 0.2$, and most of the ejecta has $v/c<0.5$. The qualitative features of the velocity distribution appear fairly robust when we vary the time at which
we measure $u_t$, and the resolution of the simulation --- with more uncertainties for the high-velocity tail of the distribution. The merger with the more compact neutron stars has a 
very similar velocity profile.
The situation is quite different for $R14i0$, where about half of the ejecta initially appear to have $v/c>0.5$. At this point, however, we do not have the ability to follow such material for a long
enough period of time to assess the reliability of the velocity estimates of that last configuration, and have to limit ourselves to the observation that the ejecta appears more relativistic for the 
largest neutron star than in the other cases studied here.

Finally, from these measurements, we can estimate the kinetic energy of the ejecta, which would be available for future emission as it slows down in the interstellar medium.
We should note that these results are only order of magnitude estimates, as these energies are sensitive to the high-velocity tail of the velocity distribution, 
which is poorly constrained in our simulations. Additionally, our energy estimates are particularly unreliable for simulation R14i0, due to the large amount of poorly resolved 
high-velocity material that is rapidly leaving the grid. We find $E_{ej}\sim (1,4,40) \times 10^{51}\,{\rm ergs}$ for simulations (R12i0,R13i0,R14i0) respectively.

Even considering the large uncertainties in these measurements, it is interesting that we consistently find that in this region of the parameter space a few percents of a solar mass
can be ejected from the system. This is indeed very different from the results obtained in the limit of low mass, low spin black holes (or for BNS~\cite{2012arXiv1212.0905H}), where only a negligible amount
of material was found to be unbound. These results indicate that in the case of $q\sim 7$ BHNS binaries, tidal disruption of the neutron star (when it occurs) is likely to be accompanied
by the ejection of $\go 10^{-2}M_\odot$ of neutron-rich material. Such outflows are promising for optical emissions due to the radioactive decay of neutron-rich elements in the
ejecta, and the production of heavy elements resulting from r-process nucleosynthesis. 
These ejecta might even be detectable as a radio afterglow as unbound material decelerates in the interstellar medium~\cite{2012ApJ...746...48M}: the kinetic energy
available for radio emission is indeed larger than in supernova explosions. However, the luminosity of the radio afterglow heavily depends on the density of the environment,
and BHNS mergers are likely to occur in much lower density environments than supernova explosions. The deceleration of the ejecta in the interstellar medium would then
occur over longer timescales, and remain harder to detect. 
Massive ejecta in BHNS mergers are also limited to high spin configurations, and should not be considered as the norm unless $\chi_{\rm BH}\sim 0.9$ is standard for black holes in 
compact binaries. And the most energetic ejecta found here,
in particular, is a very optimistic scenario for which, in addition of a high black hole spin, we used a very large neutron star. Finally, we should note that 
the detailed evolution of the tidal tail
is likely to depend on details of the equation of state that our simple $\Gamma$-law model cannot capture. 
This problem should thus be revisited carefully with a more realistic modeling of the neutron star fluid.

\subsubsection{Gravitational Waveforms}

Variations of the gravitational waveform emitted by BHNS mergers with the equation of state of the neutron star are mainly
due to two effects: the small tidal distortion of the neutron star during the inspiral, which we discussed in
Sec.~\ref{sec:tides}, and the cutoff in the gravitational-wave signal when the neutron star is disrupted (if tidal disruption
occurs). In this section, we discuss measurements of these effects in our numerical simulations, while in the next section
we will focus on their detectability by the Advanced LIGO detector.

\begin{figure}
\includegraphics[width=8.2cm]{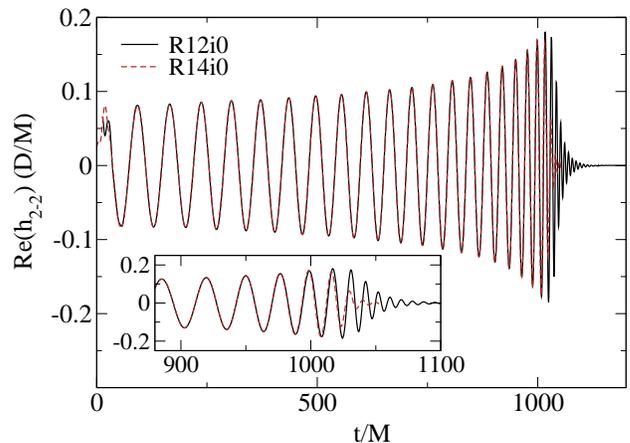}
\caption{Gravitational-wave emission [(2,-2) mode] for the non-precessing simulations R12i0 and R14i0, extrapolated to infinity
and normalized by the ratio $D/M$ of the distance to the observer $D$ to the total mass of the binary $M$.}
\label{fig:gwtimenp}
\end{figure}

\begin{figure}
\includegraphics[width=8.2cm]{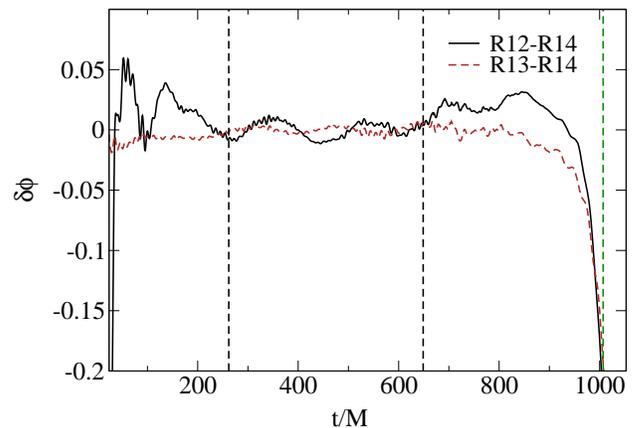}
\caption{Phase difference between the (2,-2) mode of the gravitational-wave emission of the non-precessing simulations. Both R12i0 and R13i0
are compared with R14i0. The dashed black lines show the matching interval, while the dashed green line shows the time at which the amplitude
of the signal peaks for case R14i0 (the case in which the neutron star disrupts at the earliest time).}
\label{fig:gwphasenp}
\end{figure}

The effects of tides on the gravitational-wave signal of a BHNS binary before the disruption of the neutron star are
expected to be fairly small: Damour et al.~\cite{2012PhRvD..85l3007D} computed the phase difference $\delta \Psi$
due to tidal effects
in the Fourier transform of the dominant [$(2,-2)$] mode of the waveform to 2.5PN order 
in the stationary phase approximation,
\beq
\delta \Psi^{2.5PN}_{T}=\frac{117 \tilde{\lambda}}{8\nu} x^{5/2} \hat{\Psi}^{2.5PN}_T
\label{eq:PsiPN}
\eeq
where $x=(M\omega_{\rm GW} /2)^{2/3}$ is the standard PN parameter, $\omega_{\rm GW}$ is the frequency
of the $(2,-2)$ mode of the gravitational strain,
\beq
\hat{\Psi}^{2.5PN}_T = 1 + 2.5x - \pi x^{3/2} + 8.51 x^2 - 3.92\pi x^{5/2}
\eeq
and $\tilde \lambda$ is the tidal deformability parameter which for a BHNS binary is
\beq
\tilde \lambda = \frac{1+12q}{26} \frac{2 k_2}{3C^5_{\rm NS}(1+q)^5}.
\eeq
We note that the sign of Eq.~(\ref{eq:PsiPN}) is different from the one given in~\cite{2012PhRvD..85l3007D},
due to differences in the convention used to take the Fourier transform of the signal (we use
$\tilde{h}(f)=\int h(t) e^{-i2\pi ft}$, while the dephasing $\delta \Psi$ in the stationary phase approximation
was derived with the opposite convention $\tilde{h}(f)=\int h(t) e^{i2\pi ft}$~\cite{1994PhRvD..49.2658C}).
For $\omega_{\rm GW}M\lo 0.2$ and the binary parameters considered here, we get
$\delta \Psi \lo 0.16\,{\rm rad}$ between simulations R12i0 and R14i0
(or, for the phase $\phi$ of the gravitational waveform in the time domain,
$\delta \phi \lo 0.21\,{\rm rad}$).
However, these predictions have only been tested on BNS mergers up to $x\sim C_{\rm NS}$, while the tidal
disruption of the neutron star in the binaries considered here occurs at $x\sim 0.3-0.4$. In that regime,
the PN expansion no longer appears convergent: the 2.5PN term is of the same order as
the leading order term. 


Measuring the small phase difference $\delta \phi$ in our simulation is a challenging problem. 
Fig.\ref{fig:gwtimenp} shows the dominant (2,-2) mode of the gravitational strain $h(t)$ for simulations R12i0 and R14i0, after
the application of a time and a phase shift chosen to minimize the phase difference within a matching interval
spanning 2 cycles of the radial oscillation frequency. 
Clearly, the waveforms are very similar up to the last cycle before the disruption of the larger neutron star.
Fig.~\ref{fig:gwphasenp} shows the phase differences between the waveforms of simulations R12i0, R13i0 and R14i0.
They remain under $0.05\,{\rm rad}$  until $50M$ before the peak of the gravitational-wave signal!
Moreover, the differences appear dominated by the influence of the residual eccentricity, not by equation 
of state effects. In Sec.~\ref{sec:tides}, we showed that the tidal distortion of the neutron star follows fairly closely PN
predictions, yet this is clearly insufficient to have a measurable effect on the gravitational-wave signal for most of the evolution.

\begin{figure}
\includegraphics[width=8.2cm]{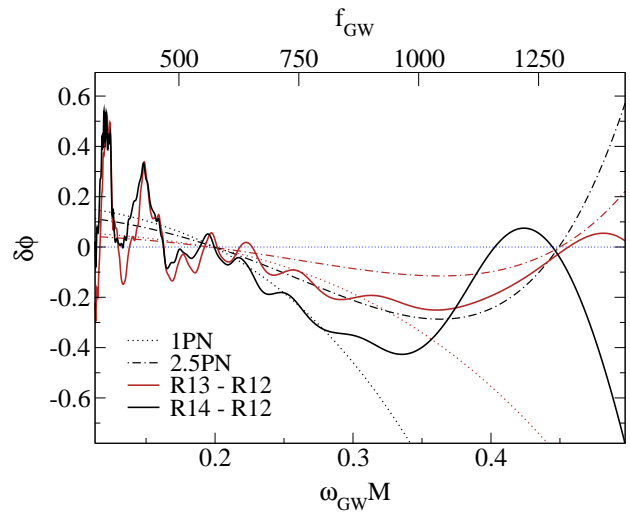}
\caption{Phase difference between the (2,-2) mode of the gravitational-wave emission of the non-precessing simulations as
a function of the normalized gravitational-wave frequency $\omega_{\rm GW}M$. The dotted and dot-dashed curves correspond to the
1PN and 2.5PN predictions from Ref.~\cite{2012PhRvD..85l3007D}. All curves are matched at $\omega_{\rm GW}M=0.2$.
At lower frequency, residual eccentricity in the simulation makes measurements of $\phi(\omega_{\rm GW})$ too noisy
to be useful.}
\label{fig:gwphiomega}
\end{figure}

A small phase difference between the waveforms can however be hidden by the matching procedure used. Another way 
to compare the waveforms is to look at the phase $\phi$ as a function of the gravitational-wave frequency. This gets
rid of the need to apply an arbitrary time shift to the waveforms. However, the computation of $\omega_{\rm GW}$ from
the numerical waveform is fairly inaccurate, and even a small residual eccentricity can introduce a large noise
in the resulting $\phi(\omega_{\rm GW})$. Computing the difference $\delta \phi(\omega_{\rm GW})$ between two numerical
simulations can thus only be done once the evolution of the orbital frequency due to orbital decay becomes fast enough to dominate the 
effects of eccentricity. Fig.~\ref{fig:gwphiomega} shows measurements of $\delta \phi(\omega_{\rm GW})$ for our numerical simulations.
For $\omega_{\rm GW}M \lo 0.2$, there are large oscillations due to the eccentricity of the orbit, and we cannot accurately measure
$\delta \phi(\omega_{\rm GW})$. But in the frequency range $0.2 \lo \omega_{\rm GW}M \lo 0.35$,
a phase shift of $\sim 0.2\,{\rm rad}$ is clearly observed between simulations R12i0 and R13i0, and the same between R13i0 and R14i0.
This result lies in between the predictions of the 1PN and 2.5PN approximations. As the 2.5PN results are barely outside of the 
expected numerical error, these simulations are not accurate enough to improve on the PN predictions for tidal effects at high frequency.
But we can confirm that the dephasing obtained from the 2.5PN predictions is at least correct within a factor of $\sim 2$ 
up to frequencies $\omega_{\rm GW}M \sim 0.35$ ($f\sim1\,{\rm kHz}$). The 1PN prediction also seem slightly more accurate than the 2.5PN 
results.

\begin{figure}
\includegraphics[width=8.2cm]{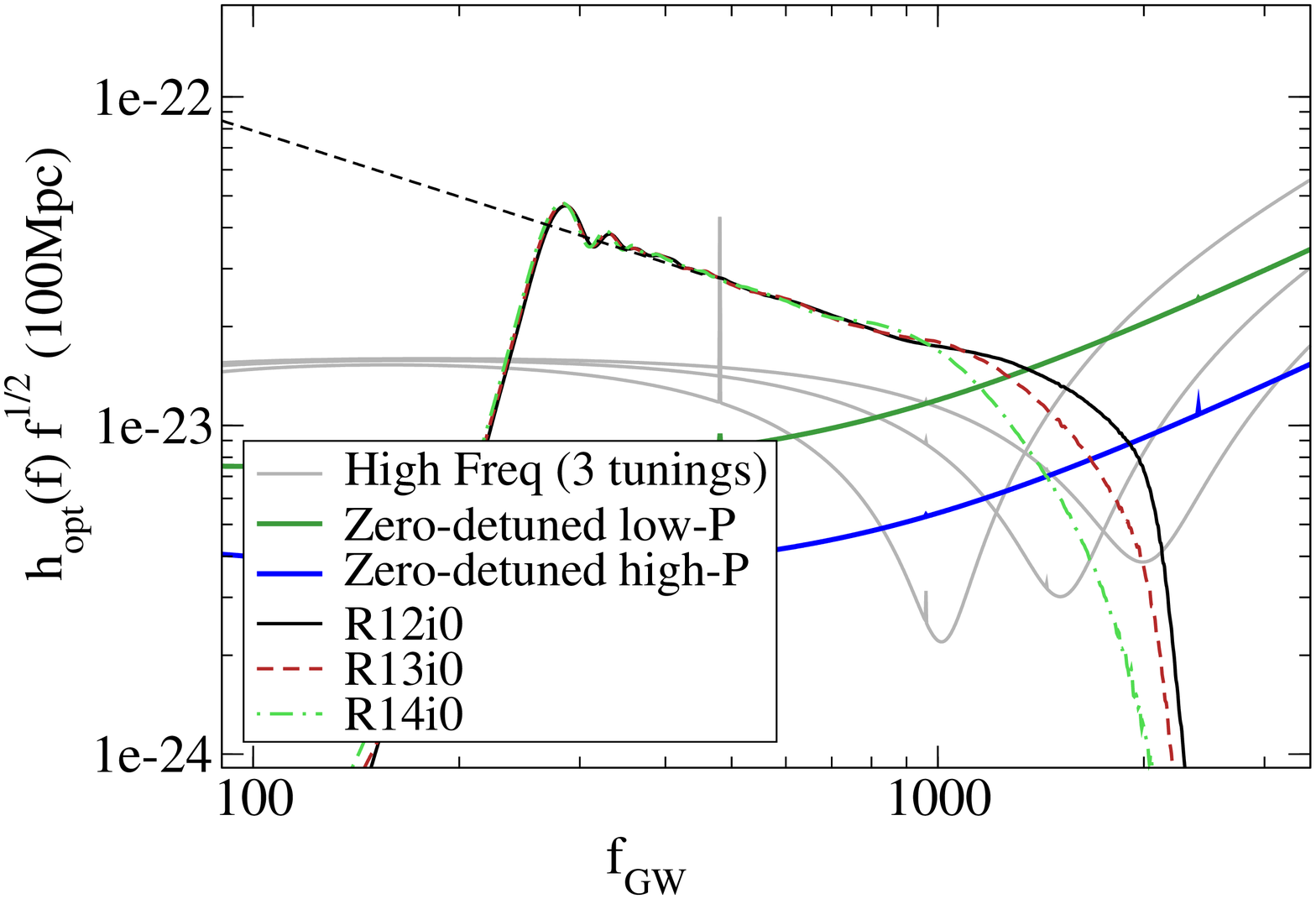}
\caption{Spectrum of the gravitational-wave signal. $h_{\rm opt}(f)$ is the spectrum of the dominant mode of the gravitational-wave signal as seen by an optimally oriented observer at $100\,{\rm Mpc}$. 
The dashed line shows the leading order PN behavior, $h=A f^{-7/6}$, with amplitude $A$ matched to the numerical results. The Zero-Detuned High Power and Zero-Detuned Low Power noise curves of Advanced
LIGO~\cite{Shoemaker2009} are also shown, together with 3 potential high-frequency tunings of
the detector, at $1\,{\rm kHz}$~\cite{Shoemaker2009}, $1.5\,{\rm kHz}$ and $2\,{\rm kHz}$~\cite{Ligo_HFcurve}.}
\label{fig:gwspecnp}
\end{figure}

The effects of the disruption of the neutron star by the tidal field of the black hole are easier to see in the numerical results.
Fig.~\ref{fig:gwspecnp} shows the spectrum of the gravitational-wave signal for the three simulations R12i0, R13i0 and R14i0 (for
an optimally oriented binary at $100\,{\rm Mpc}$).
The largest star disrupts earlier, and the gravitational-wave spectrum is cut slightly above $1\,{\rm kHz}$. For the smallest
star, the cutoff is at about $2\,{\rm kHz}$. As opposed to tidal effects, this high-frequency cutoff is very well resolved in
the simulations. 

\begin{table}
\caption{
Gravitational-wave emission over the course of the simulation as measured
at a radius $R=275M$, where $M$ is the ADM mass of the system at infinite
separation. $E^{\rm GW}$ is the energy contained in the waves, $J^{\rm GW}$
their angular momentum, and $v_{\rm kick}=P^{\rm GW}/M_{\rm BH}^{\rm final}$ the
velocity kick given to that black hole. $f_{\rm cut}^{\rm GW}$ is the cutoff
frequency of the gravitational-wave signal defined by Eq.~\ref{eq:fcut}. 
}
\label{tab:gw}
\begin{tabular}{|c||c|c|c|c|}
\hline
Name & $E^{GW}/M$ & $J^{GW}/M^2$ & $v_{\rm kick}$(km/s) & $f_{\rm cut}^{\rm GW}$ (kHz)\\ 
\hline
R12i0 & 0.021 & 0.16 & 30 & 2.1 \\ 
R13i0 & 0.017 & 0.15 & 45 & 1.8 \\ 
R14i0 & 0.014 & 0.13 & 45 & 1.5 \\ 
R14i20 & 0.013 & 0.13 & 60 & --- \\ 
R14i40 & 0.013 & 0.11 & 150 & --- \\ 
R14i60 & 0.013 & 0.09 & 345 & --- \\ 
\hline
\end{tabular}
\end{table}

Other physical quantities which can be extracted from the gravitational-wave signal are summarized in Table~\ref{tab:gw}: 
the energy and angular momentum radiated, the final kick given to the system, and a cutoff frequency $f_{\rm cut}$ defined (arbitrarily) by the relation
\beq
\label{eq:fcut}
2h(f_{\rm cut})f_{\rm cut}^{7/6}= h(f_{\rm ref})f_{\rm ref}^{7/6}
\eeq
with $f_{\rm ref}=0.5\,{\rm kHz}$ (Note that any value of $f_{\rm ref}$ in the range $0.3\,{\rm kHz}-0.8\,{\rm kHz}$ gives nearly 
identical result as $hf^{7/6}$ is approximately constant during the inspiral,
as shown in Fig.~\ref{fig:gwspecnp}).
As the binding energy of these systems at $t=0$ is $E^{\rm bind}_0=0.0055M$, we see that these binaries will radiate $\sim 2\%-2.5\%$ of their energy
before merging, and $\sim 15\%$ of their angular momentum (the more compact neutron stars naturally radiating more, as they disrupt later). 
The final kicks remain low ($<30\,{\rm km/s}-50\,{\rm km/s}$), as is generally observed for non-precessing BHNS binaries.

\subsubsection{Detectability of the Neutron Star Radius by Advanced LIGO}
\label{sec:npgw}

Keeping in mind the results of the previous section, we can begin to address another important question: the measurability of finite size
effects on the gravitational waveform of BHNS mergers for mass ratios $q\sim 7$. An earlier analysis
of these issues by Lackey et al.~\cite{2012PhRvD..85d4061L} showed that at lower mass ratios ($q=2-3$) and for nonspinning black holes 
Advanced LIGO would be sensitive to differences in the radius of the neutron star of order $10\%-40\%$ for an optimally 
oriented BHNS merger located at $100\,{\rm Mpc}$. This is due in part to the effects on the waveform of the tidal distortion
of the neutron star, and in part to variations in the binary separation at which the neutron star disrupts and
the gravitational-wave signal is cut off.

At higher mass ratio, tidal effects are smaller. However, the disruption of the neutron star occurs at a lower frequency
and the amplitude of the gravitational-wave signal is larger. It is thus unclear whether finite size effects will be easier or harder
to detect. On Fig.~\ref{fig:gwspecnp}, we show the spectrum of the gravitational-wave signal
as seen by an optimally oriented observer located $100\,{\rm Mpc}$ away from the binary, and compare it with different Advanced LIGO detector's strain
noise spectra (see below). At that distance the differences between the three simulations seem to be marginally measurable. 
In the rest of this section, we will attempt to quantify this statement more carefully.  

To determine whether the difference
between two waveforms $h_1$ and $h_2$ can be detected by Advanced LIGO,
we use the approximate condition~\cite{Lindblom2008}  
\beq
\|\delta h\|^2 = \langle \delta h,\delta h \rangle = \langle h_1-h_2,h_1-h_2 \rangle \geq 1.
\label{eq:dh}
\eeq
where the inner product is defined as
\beq
\langle g,h \rangle =  2 \int_0^{\infty} df
\frac{\tilde{g}^*(f)\tilde{h}(f) + \tilde{g}(f)\tilde{h}^*(f)}{S_n(f)}.
\label{eq:gdoth}
\eeq
Here $\tilde{g}(f)$ and $\tilde{h}(f)$ are the Fourier transforms of
two waveforms $g(t)$ and $h(t)$, and $S_n(f)$ is the one-sided power
spectral density of the detector's strain noise, defined as 
\beq
S_n(f) = 2 \int_{-\infty}^{\infty} d \tau \, e^{2 \pi i f \tau}\,
C_n(\tau)\;,\qquad f>0,
\eeq
where $C_n(\tau)$ is the noise correlation matrix for zero-mean,
stationary noise. In this case, we will consider three of the Advanced LIGO guideline noise curves
defined in Ref.~\cite{Shoemaker2009}: the Zero Detuned Low Power spectrum, which is the
expected sensitivity of the detector once signal recycling mirrors are in place, the 
Zero Detuned High Power spectrum, which is the final design sensitivity of Advanced LIGO,
and a High Frequency noise curve optimized to take data at $1\,{\rm kHz}$.
We also consider alternative tunings of the Advanced LIGO detector to $1.5\,{\rm kHz}$
and $2\,{\rm kHz}$, using noise curves graciously provided to us by Nicolas Smith-Lefebvre~\cite{Ligo_HFcurve},
and generated by the noise simulation package `gwinc' developed by the LIGO collaboration.
These noise curves assume an Advanced LIGO detector in the same configuration as the High Frequency
model of Ref.~\cite{Shoemaker2009}, except that the signal-recycling mirror detuning is chosen to tune the
detector to higher frequencies.
 
When taking the inner product $\|\delta h\|^2$, we choose one polarization of $h_1$
and then allow for a time and phase shift in $h_2$, chosen to maximize the inner
product $\langle h_1,h_2 \rangle$ (which is identical to minimizing $\|\delta h\|^2$ for $\|\delta h\|^2 \ll \|h_{1,2}\|^2$).
Finally, we consider only the quadrupolar part of the waveform as measured by an optimally oriented observer:
\beq
h_{\rm opt}(t) = \sqrt{\frac{5}{4\pi}}h_{2,2}(t).
\eeq

In theory, the integration in Eq.~(\ref{eq:gdoth}) should be carried
over the entire frequency band of the detector. This is however complicated for our waveforms, as the numerical simulations
only cover the high-frequency part of the signal ($f>0.25\,{\rm kHz}$). To construct a full waveform, the numerical results
should be hybridized with some analytical approximation valid at low frequency (PN, Effective One-Body,...). 
But in the region of parameter space considered here ($q=7$, $\chi_{\rm BH}=0.9$), the error
coming from extending these approximants to  
frequencies $f\sim 0.25\,{\rm kHz}$ is significantly larger than the
actual differences expected between waveforms for $f<0.25\,{\rm kHz}$. 
Uncertainties in the construction of the hybrid then dominate the
measurement of $\|\delta h\|^2$. 

We will instead consider three approximations to $\|\delta h\|^2$:
\begin{itemize}
\item Limiting the integration to frequencies $f>0.8\,{\rm kHz}$, where the waveform is known exactly from the numerical simulations.
This neglects tidal effects during the inspiral.
\item Limiting the integration to frequencies $f<0.8\,{\rm kHz}$, and using PN predictions over the entire frequency range (see Appendix~\ref{app:match}). 
In this case, we ignore the effects of the disruption of the neutron star, as well as errors in the PN predictions at high frequencies. As seen in the previous
section, the 1PN predictions appear to remain fairly accurate for $f<0.8\,{\rm kHz}$, and provide at least a good qualitative estimate
of the tidal effects during the inspiral. 
\item Combining the low-frequency PN predictions with the numerical results at high frequency by matching in the frequency domain 
the phase difference between two simulations to the predicted PN phase difference. This procedure is detailed in Appendix~\ref{app:match}.
\end{itemize}
The matching procedure and the differences between various PN orders each cause uncertainties of $\sim 10\%$ in $\|\delta h\|$.

\begin{table}
\caption{
Detectability of tidal effects in non-precessing BHNS mergers
for 3 different Advanced LIGO noise curves from Ref.~\cite{Shoemaker2009}: 
Zero-Detuned High Power, Zero-Detuned Low Power and High Frequency tuned at
$1\,{\rm kHz}$ (see also Fig.~\ref{fig:gwspecnp}). 
The quantities $\|\delta h\|$, defined in Eq.~\ref{eq:dh},
are the detectability criteria for optimally oriented events at $100\,{\rm Mpc}$.
We consider first the results for the numerical waveform limited to $f>0.8\,{\rm kHz}$
(NR Only), then for the first order PN expansion [Eq.~(\ref{eq:PsiPN})] limited
to $f<0.8\,{\rm kHz}$ (1PN Only) and finally for hybrid waveforms matched
in spectral space between $0.3\,{\rm kHz}<f<0.8\,{\rm kHz}$ (Hybrid-1PN).
For each case, we compare simulations [R12i0,R14i0] ($||\delta h||_{\rm 12-14}$),
[R12i0,R13i0] ($||\delta h||_{\rm 12-13}$) and [R13i0,R14i0] ($||\delta h||_{\rm 13-14}$).
For the hybrids, we also give results for noise curves tuned to $1.5\,{\rm kHz}$ and
$2\,{\rm kHz}$.
}
\label{tab:diff}
{\bf NR Only}\\
\begin{tabular}{|c|c|c|c|}
\hline
$S(f)$ & $||\delta h||_{\rm 12-14}$ & $||\delta h||_{\rm 12-13}$ & $||\delta h||_{\rm 13-14}$\\ 
\hline
Zero Det H-P & 1.5 & 0.7 & 0.9 \\
Zero Det L-P & 0.7 & 0.3 & 0.4 \\
High Freq($1\,{\rm kHz}$)   & 1.2 & 0.4 & 0.9 \\
\hline
\end{tabular}
\\
{\bf 1PN Only}\\
\begin{tabular}{|c|c|c|c|}
\hline
$S(f)$ & $||\delta h||_{\rm 12-14}$ & $||\delta h||_{\rm 12-13}$ & $||\delta h||_{\rm 13-14}$\\
\hline
Zero Det H-P & 1.0 & 0.4 & 0.6 \\
Zero Det L-P & 0.5 & 0.2 & 0.3 \\
High Freq($1\,{\rm kHz}$)   & 0.4 & 0.2 & 0.3 \\
\hline
\end{tabular}
\\
{\bf Hybrid-1PN}\\
\begin{tabular}{|c|c|c|c|}
\hline
$S(f)$ & $||\delta h||_{\rm 12-14}$ & $||\delta h||_{\rm 12-13}$ & $||\delta h||_{\rm 13-14}$\\
\hline
Zero Det H-P & 2.5 & 1.2 & 1.5 \\
Zero Det L-P & 1.2 & 0.6 & 0.7 \\
High Freq($1\,{\rm kHz}$)   & 1.5 & 0.6 & 1.0 \\
High Freq($1.5\,{\rm kHz}$) & 3.2 & 1.4 & 1.9 \\
High Freq($2.0\,{\rm kHz}$) & 2.6 & 1.3 & 1.4 \\
\hline
\end{tabular}
\end{table}

The results for each method are summarized in Table~\ref{tab:diff}. We find that for the Zero Detuned noise curves,
the high frequency cutoff is only $\sim 50\%$ more important than the low-frequency tidal effects. The tuned High Frequency
noise curves, quite naturally, are much more sensitive to the disruption of the neutron star than to tidal effects.
For the Zero Detuned High Power noise curve, and using our estimates of the mismatch $||\delta h||$ over the entire
LIGO band, the detectability criteria is satisfied if the neutron stars have
radii differing by
\beq
\Delta R_{\rm NS} \go \frac{D}{125\,{\rm Mpc}} \,{\rm km},
\eeq 
where $D$ is the distance to the observer.
The Low Power noise curve requires the binary to be about twice as close. 
However, using the low-power detector with high frequency tuning (at $\sim1.5\,{\rm kHz}$) leads
to $||\delta h||$ larger than for the Zero Detuned High Power noise curve by $\sim 20\%-30\%$.

Differentiating a binary black hole system from a BHNS binary would only be slightly easier. The phase difference between a
point-particle waveform and the waveform of simulation R14i0 is, during inspiral, only $\sim 1.5$ times larger than the 
phase difference between simulations R12i0 and R14i0. And the high-frequency cutoff in the waveform would not help us much,
as the most compact star considered here (R12i0) does not disrupt very far from the ISCO.

These results are not very promising. For the Zero Detuned High Power noise curve 
and the waveforms from simulations R12i0 and R14i0, the condition $||\delta h||>1$ 
will only be satisfied for binaries with signal-to-noise ratio $\rho\go 26$,
or about $3\%$ of the Advanced LIGO events (assuming that BHNS binaries are equally distributed in volume).
A detector tuned at $1.5\,{\rm kHz}$ would reach the condition $||\delta h||>1$ for about twice as many events.
The criteria $||\delta h||>1$ is also an optimistic limit: it does not take into
account the fact that all other parameters of the binary are here assumed to be known. Uncertainties in the masses
of the objects and the spin of the black hole will significantly affect these results, making it harder to detect
equation of state effects. Additionally, both the PN dephasing and the variations in the high-frequency
cutoff of the signal are helped by the fact
that we are considering a rapidly rotating black hole. For a nonspinning black hole, the neutron star would reach the ISCO at
lower frequencies, and tidal effects during the inspiral would be even smaller. And as the neutron star would plunge into
the black hole before being disrupted, there is no guarantee that there would be a measurable difference in the cutoff frequency
of the waveform (although this question certainly deserves further investigation: the merger would also occur in a more
favorable frequency range). Finally, any real data analysis would require the knowledge of the 
waveform at low frequency, which is not at this point known with enough accuracy for binaries with $q=7$ and $\chi_{\rm BH}=0.9$.
The theoretical detectability conditions considered here are thus certainly too generous.

\subsection{Precessing Binaries}

Our second set of BHNS mergers considers variations of the orientation of the black hole spin. The starting configuration
is the largest neutron star studied in the previous section. The black hole spin is inclined with respect to the orbital angular momentum by
$\theta_{\rm BH}=20^\circ,40^\circ,60^\circ$. In all cases, the misaligned component of the spin is initially along the line connecting the
black hole and neutron star centers. We do not expect this choice to affect the qualitative feature of the 
merger (disruption, disk formation)~\cite{2011PhRvD..83b4005F}. 
It should, however, influence the magnitude of the velocity kick given to the
final black hole as a result of gravitational-wave emission~\cite{2009PhRvD..79f4018L}.
Modifying the initial separation of the binary is effectively identical to a change in the orientation of the black hole spin at constant
$\theta_{\rm BH}$. Indeed, for a binary in which only one object is spinning, $\theta_{\rm BH}$ is conserved~\cite{1994PhRvD..49.6274A} 
and changing the initial separation only modifies the phase of the precession of the binary.

\subsubsection{Inspiral: Orbital Precession}

\begin{figure}
\includegraphics[width=8.2cm]{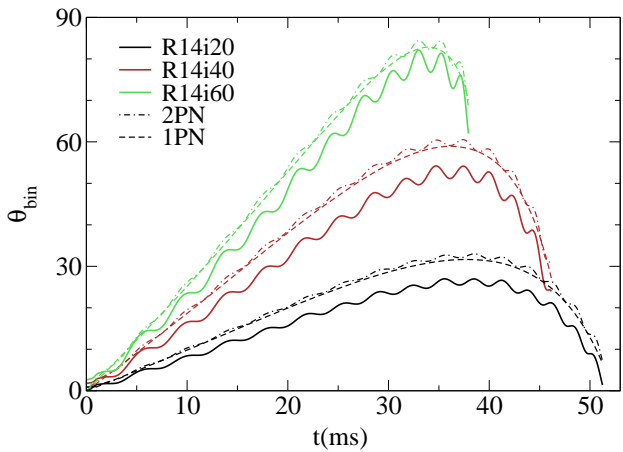}
\caption{Inclination $\theta_{\rm bin}$ between the initial and current direction of the normal
to the orbital plane.}
\label{fig:orbitprec}
\end{figure}

During inspiral, the main difference with the aligned configurations will naturally be the precession of both the black hole spin and the orbital
angular momentum around the total angular momentum of the system. A simple coordinate measurement of that precession is presented in
Fig.~\ref{fig:orbitprec}, using the angle $\theta_{\rm bin}$ between the direction of the initial orbital angular momentum and the 
normal to the orbital plane (defined as $(c_{\rm BH}-c_{\rm NS})\times (v_{\rm BH}-v_{\rm NS})$, where
$(c_{\rm NS},c_{\rm BH})$ are the coordinates of the centers of the compact objects, and 
$(v_{\rm BH},v_{\rm NS})$ their velocities).

We see that the amplitude of the precession of the orbital plane vary from $27^\circ$ for R14i20 to $82^\circ$ for the
most inclined case. If the total angular momentum of the system was constant (i.e. in the absence of gravitational-wave emission), we would
expect the amplitude of that precession to be twice the initial angle between the orbital angular momentum and the total angular momentum,
i.e. $25^\circ$ and $78^\circ$ for simulations R14i20 and R14i60 respectively. Somewhat larger values are expected for radiating systems, as
the loss of angular momentum due to gravitational-wave emission is to first order aligned with the current orbital angular momentum of the binary:
taking into account the angular momentum lost to gravitational waves listed in Table~\ref{tab:gw} would correct these estimates to $30^\circ$ 
and $85^\circ$ (which are now overestimates, as part of the radiated angular momentum is emitted during the disruption of the neutron star).
Over the course of the binary evolution, the most inclined binary goes through slightly more than half of a precession period, while R14i20 gets close
to completing a full precession period (as $\theta_{\rm bin}$ is defined with respect to the initial orbital plane, we get $\theta_{\rm bin}\approx 0$
after a full precession period, and not after half a precession period).

A simple comparison of the measured precession of the orbital plane with the PN predictions from~\cite{2006PhRvD..74j4033F} 
is also presented on Fig.~\ref{fig:orbitprec}. The 1PN and 2PN curves are obtained by evolving the initial BH spin and orbital angular momentum 
using the PN formulae from~\cite{2006PhRvD..74j4033F}, but assuming that the trajectories of the compact objects are those observed 
in the simulation (i.e. when computing the relative position and velocity of the objects, we use our results and not a PN evolution of the 
initial conditions, as the PN equations of motion are quite inaccurate so close to merger). 
We see that the period over which the orbital plane precesses in the simulations matches 
the PN predictions very well, while the amplitude of the precession is $\sim 10\%$ smaller in the numerical results. Such differences (as well as
the additional oscillations in the value of the inclination angle) could however easily be due to the fact that these measurements are certainly not
gauge-independent. 

A large precession of the orbital plane is quite natural for the high mass ratio, high spin systems considered here. Indeed, the angular momentum 
of the black hole is $J_{\rm BH}/M^2 = 0.689$ while the initial orbital angular momentum is only $L_{\rm orbit}/M^2 = 0.354$ (and about $10\%$ of the
total angular momentum, or $\sim 30\%$ of $L_{\rm orbit}$, is radiated in gravitational waves). This means that even for relatively low 
inclinations of the black hole spin (such as in simulation R14i20), large variations of the orientation of the orbital plane occur.
These oscillations and, more importantly, the shift in the phase of the gravitational waveform with respect to aligned spin templates which accompany
them, can make the detection of precessing binaries challenging. The higher dimensionality of the parameter space to consider also complicates
parameter estimates --- although there are also positive effects due to precession: some of the degeneracies existing between the parameters 
of the system for aligned binaries are broken for precessing systems~\cite{2012PhRvD..85l3007D}. 
As mentioned in the previous section, an inaccurate determination of the
parameters of the system increases the error in any measurement of the neutron star equation of state from gravitational waveforms. 
Obtaining  proper constraints on the parameters of a BHNS binary with misaligned spin, which requires reliable templates for precessing systems, 
is thus a prerequisite to any attempt at constraining the equation of state of neutron stars from BHNS waveforms, at least for the high-spin configurations
considered here (unless the spin of the black hole is aligned with the orbital angular momentum of the system by some unknown mechanism
during the pre-merger evolution of the binary).

\subsubsection{Disruption and Disk Formation}

\begin{figure*}
\includegraphics[width=6cm]{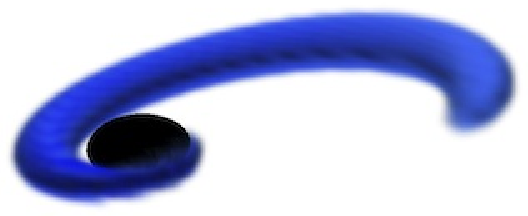}
\includegraphics[width=6cm]{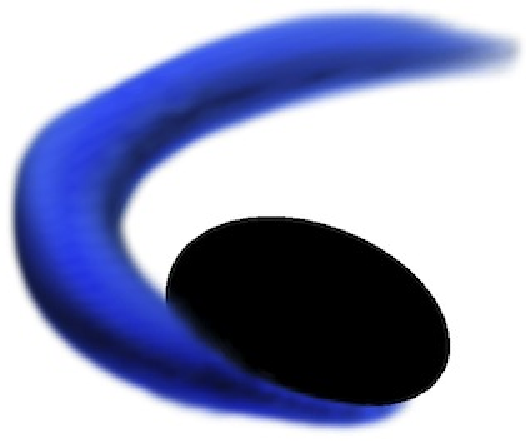}\\
\includegraphics[width=6cm]{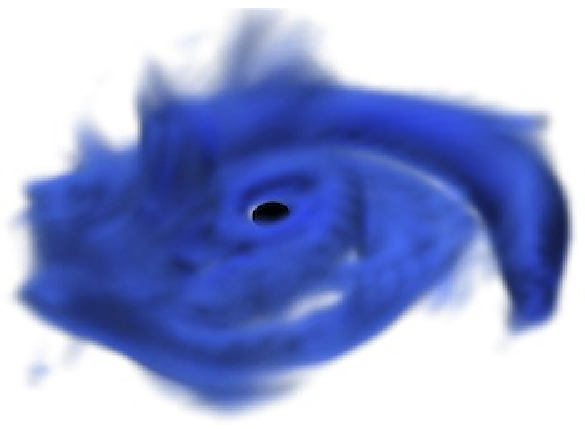}
\includegraphics[width=6cm]{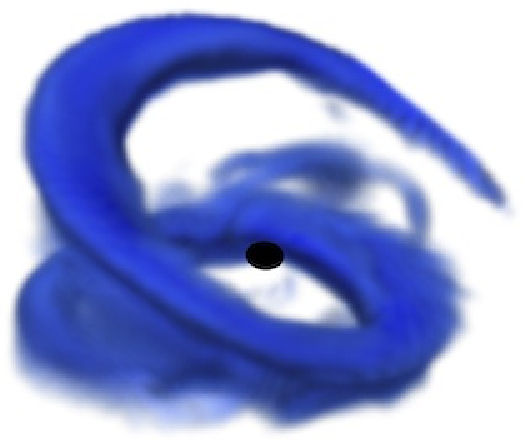}
\caption{Same as Fig.~\ref{fig:simnp}, but for the precessing binaries R14i20 (left) and R14i40 (right).}
\label{fig:simp}
\end{figure*}

The effect of spin misalignment on the properties of the remnant of a BHNS merger were first studied in a general relativistic framework
in Ref.~\cite{2011PhRvD..83b4005F}. Material
on an orbit inclined with respect to the equatorial plane of the black hole reaches the region in which stable orbits no longer exist at a larger
separation than material in the equatorial plane on a corotating orbit. Effectively, this means that BHNS mergers with misaligned black hole
spins are roughly equivalent to mergers with a lower black hole spin aligned with the orbital angular momentum.
Disruption becomes less likely for misaligned configurations, and the mass remaining outside of the black hole
at late times is smaller. The smallest radius at which stable orbits exist for black holes with $\chi_{\rm BH}=0.9$ at inclination 
$\theta_{\rm BH}=(20^\circ,40^\circ,60^\circ)$ is equal to the radius of the innermost stable circular orbit of a black hole with aligned spin 
$\chi_{\rm BH}=(0.89,0.80,0.62)$~\cite{2001PhRvD..64f4004H,2007ApJ...668..417F}.
\footnote{The method used here to approximate the ``effective'' spin
of a BHNS binary for the purpose of tidal disruption was first pointed out to us by Nicholas Stone. This method has been used to impose constraints on systems
which could lead to SGRBs in Stone et al.~\cite{2012arXiv1209.4097S}. 
Our numerical simulations tend to confirm that this is indeed a reasonable approximation to the result
of tidal disruption in precessing BHNS binaries.}
And indeed, the disruption of the neutron star is close to what is expected for such
spins: for the low inclination case R14i20, the mass of material remaining outside of the black hole after merger is nearly identical to what was
observed in the aligned configuration R14i0 while in simulations R12i40 and R14i60, $15\%$ and $3\%$ of the neutron star mass remain outside of 
the hole $5\,{\rm ms}$ after merger. This is very similar to what our simple fitting model~\cite{2012arXiv1207.6304F} 
would predict ($13\%$ and $1\%$ of the neutron star matter surviving the merger for $\chi_{\rm BH}=(0.80,0.62)$), 
or what might be inferred from numerical simulations for smaller black hole spins~\cite{2012PhRvD..85d4015F} (which found $6\%$ of the matter remaining outside
of the hole for $\chi_{\rm BH}=0.7$ and none for $\chi_{\rm BH}=0.5$, all other parameters being identical to the cases considered here). It thus seems
fairly likely that, for the purpose of measuring the mass of material remaining outside of the black hole at late times at least, 
the disruption of the neutron star in BHNS mergers 
with misaligned black hole spins can be modeled with good accuracy by considering the results of aligned configurations only.

However, there are important qualitative differences between the behavior of aligned and misaligned configurations. 
Fig.~\ref{fig:simp} shows snapshot of the two simulations with the lowest inclination angle for the black
hole spin (R14i20 and R14i40), at a time at which $50\%$ of the neutron star has been accreted onto the black hole ({\it top}) as well as $5\,{\rm ms}$
later ({\it bottom}). These can be directly compared with Fig.~\ref{fig:simnp} for non-precessing systems. Simulation R14i20 mostly behaves as the aligned
case, although the slight precession of the tidal tail with respect to the disk induces small differences in the formation of the disk, and should affect
its subsequent evolution as matter falls back from changing directions. At moderate inclinations (R14i40), the changes are more drastic.
There is not much of a disk forming: most of the remaining material is in a long tidal tail, which is differentially precessing. For aligned configurations,
a disk generally starts to form as the front edge of the accretion flow wraps around the black hole and hits the material from the tidal tail, creating
a shock, outflows, and a rapid redistribution of the tidal tail material. Disk-tail interactions are much less visible in inclined simulations (although there
are still contacts between the inner and outer edge of the tidal tail as it wraps around the black hole). Existing shocks are however
still sufficient to heat the remaining material in simulations R14i20 and R14i40 to an average temperature $<T>\sim 2\,{\rm MeV}-3\,{\rm MeV}$. Not surprisingly 
the remnant of the most inclined merger (R14i60), which does not form a disk, is much cooler ($T<1\,{\rm MeV}$).
Finally, we find that precession does not prevent the formation of a baryon-poor region along the rotation axis of the black hole: a cone of opening
angle $\Theta_{\rm clean}\go 30^\circ$ is clear of material at densities $\rho_0>10^{9}\,{\rm g/cm^3}$, at least over the short timescale over which the post-merger remnant is evolved. 
Close to the black hole, lower density material is not resolved in the simulation. This baryon-poor region was also shown to exist after the merger
of precessing BHNS binaries at lower mass ratio~\cite{2011PhRvD..83b4005F}.

The significant asymmetry of this system with respect
to the equatorial plane of the black hole, as well as the differential precession between fluid elements, could also have important consequences for the long
term evolution of the system. 
Disk simulations by McKinney et al.~\cite{2013Sci...339...49M} indicate that coupling between the magnetic field in an accretion disk (and relativistic jet) 
and the spin of the 
black hole leads to an alignment of the inner disk with the spin of the black hole, while the outer disk remains misaligned. Similarly,
the jet is emitted along the rotation axis of the black hole, but orthogonal to the plane of the outer disk at large distances. How these
effects will play out for the more compact disk produced by BHNS mergers is an open question.
Another important consequence of spin-orbit misalignment was pointed out by Etienne et al.~\cite{2012PhRvD..86h4026E}, who showed
that asymmetries and motion across the equatorial plane of the black hole contribute to the formation of a larger toroidal field within 
the disk formed by BHNS mergers (aligned configurations with equatorial
symmetry, on the other hand, form disks with mostly poloidal magnetic fields, as the magnetic field comes from the winding of the original field lines frozen
within the disrupting neutron star). The toroidal field is amplified by the fastest growing mode of the magnetorotational instability(MRI), the
mechanism most likely to allow the magnetic field in the accretion disk to grow up to levels at which jets (and short gamma-ray bursts) could
be created. Misaligned black hole spins could thus lead to qualitative difference in the evolution of magnetized disks. They would also make it easier to resolve the growth
of the MRI numerically, as the wavelength of the fastest growing MRI mode scales with the magnitude of the toroidal component of the magnetic 
field~\cite{2010MNRAS.408..752P}.
On the other hand, the disks formed from precessing BHNS binaries have lower densities and lower masses. Compared to aligned configurations, a larger fraction
of the matter remaining outside of the black hole is sent in the tidal tail. The most inclined configuration studied here (R14i60), which barely disrupts,
actually keeps less than $1\%$ of the neutron star material within $200\,{\rm km}$ of the black hole. Nearly all of the remnant mass ($3\%$ of the neutron star material)
is on highly eccentric, differentially precessing orbits. 

If the velocity kicks given to neutron stars during supernova explosions are such that a majority of BHNS systems are 
within the range of inclination of the black hole spin studied here (as might be expected from the results of Belczinsky et al.~\cite{2008ApJ...682..474B}), 
the effects of inclination on the orbital evolution of the binary,
and consequently on the gravitational-wave signal, would be significant. But the conditions required for the disruption of the neutron star to occur in 
a significant number of systems would not be dramatically modified from those derived in non-precessing systems: the location of the marginally stable orbit
does not change much for inclinations $\theta_{\rm BH}<30^\circ$ (see e.g.~\cite{2007ApJ...668..417F}), at least for the relatively large spins $\chi_{\rm BH}>0.7$
which we already know are necessary for tidal disruption to occur for mass ratios $q\sim 5-10$. The remnant disks would be less massive,
but with larger toroidal magnetic fields initially. Considering that even low mass disks ($\sim 0.01 M_\odot$) are energetically sufficient to power
gamma-ray bursts if the conditions (temperature, magnetic fields) are otherwise right~\cite{2005ApJ...630L.165L,2012arXiv1210.8152G}, 
misaligned configurations could in the end prove more favorable than the aligned cases.

\subsubsection{Waveforms}

\begin{figure}
\includegraphics[width=8.2cm]{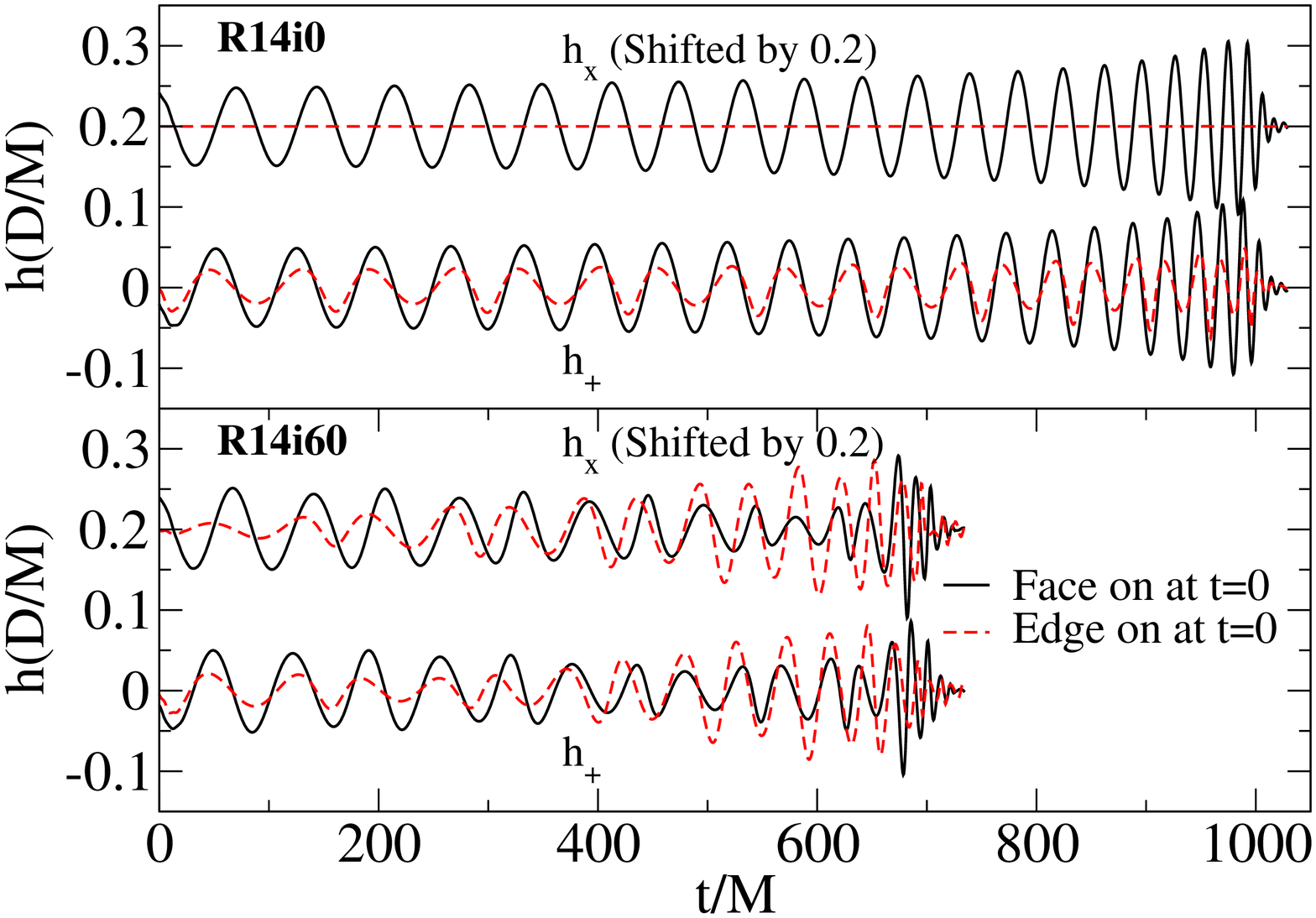}
\caption{Gravitational waveforms as measured by observers who, at $t=0$, see the binary face-on (solid black line) and edge-on along the line connecting the centers of the two objects (dashed red line). $h$ is normalized by the ratio $D/M$ of the distance to the observer $D$ to the total mass of the binary $M$. We show both the $h_+$ and $h_\times$ 
polarization, with $h_\times$ shifted by $0.2$.
{\it Top:} Non-precessing configuration R14i0. The face-on observer is always optimally located, while the edge-on observer receives a weaker signal in $h_+$ 
(in which higher order modes are however more visible), and no signal at all in
$h_\times$.
{\it Bottom:} Precessing configuration R14i60. The envelope of the signal varies in time as the binary precess, and the optimal orientation is modified accordingly.
The merger also occurs at an earlier time, as the orbital hang-up is only due to the aligned component of the black hole spin.}
\label{fig:pwaves}
\end{figure}

The waveforms from precessing BHNS binaries are significantly different from their non-precessing counterparts, as previously mentioned: the
precession of the orbital plane shown in Fig.~\ref{fig:orbitprec} will cause a modulation of the preferred direction for gravitational-wave emission (see Fig.~\ref{fig:pwaves}),
which has to be properly modeled in order to avoid significantly reducing our sensitivity to any waveform emitted by a precessing system. The modeling
of precessing waveforms goes beyond the scope of this article. A more detailed study of the impact of precession on the detectability of BHNS mergers
can be found in Brown et al.~\cite{2012PhRvD..86f4020B}. Assuming an isotropic distribution of black hole spin, about half of the BHNS binaries within the theoretical range
of the next generation of gravitational-wave detectors could be missed given the performance of the current search methods in the regime of high mass
ratio, strongly precessing systems. Attempts to reduce the complexity of the problem by studying the waveforms in a preferred frame precessing
with the binary are under way~\cite{2011PhRvD..84b4046S,2011PhRvD..84l4011B,2012arXiv1207.3088S}, and could help in the construction of future precessing templates, but the detection of
BHNS systems in Advanced LIGO remains an important challenge today (see also~\cite{2012arXiv1210.6666A} for an updated template bank for binaries
with arbitrary spins). 
Considering the negligible influence of tidal effects on the waveform before the disruption of the neutron star, these issues are however better studied in the context
of black-hole--black-hole binaries, for which longer and more accurate precessing waveforms are available. Results obtained for these systems
should be immediately applicable to BHNS inspirals, at least in the range of mass ratios considered in this work (tidal effects are more important
for more symmetric mass ratios).

An interesting particularity of BHNS mergers with high black hole spins, precession, and a high enough mass ratio that tidal disruption either does not occur
or only occurs very close to the marginally stable orbit, is the possibility for the remnant black hole to receive a significant velocity kick from the merger.
This effect is already well-known for binary black hole systems~\cite{2011CQGra..28k4015Z,2011PhRvL.107w1102L}: binary black holes with parameters similar to 
the BHNS mergers studied here
would receive velocity kicks $v_{\rm kicK} \lo 2000\,{\rm km/s}$ (and $v_{\rm kick}\lo 5000\,{\rm km/s}$ for more favorable parameters). 
Kicks in BHNS mergers are
generally much smaller, with $v_{\rm kick}\sim 50\,{\rm km/s}-100\,{\rm km/s}$, even in the precessing configurations that we previously studied~\cite{2011PhRvD..83b4005F}.
This is due to the fact that most of the kick is received at the time of merger, while in BHNS systems at lower mass ratio the neutron star disrupts 
before merger, effectively cutting off the asymmetric gravitational-wave emission responsible for the kick. Here, however, the disruption of the neutron
star occurs very late in the inspiral --- or nearly not at all in the case of the most inclined configuration. The strong distortion of the neutron star as it plunges
into the black hole will still reduce gravitational-wave emission at merger, even for configurations in which no matter remains outside of the black hole
afterwards, but not nearly as much as for aligned configurations, or for more symmetric mass ratios. Accordingly, we find that much larger kicks 
$v_{\rm kick}\sim 345\,{\rm km/s}$ are now possible. It should also be noted that the kick obtained from binary mergers is proportional to $\cos{(\phi+\phi_0)}$,
with $\phi$ the orbital phase at merger and $\phi_0$ an unknown phase shift. The only way to measure the maximum kick from a specific configuration
is thus to consider a sequence of mergers, spanning a range of phases $\phi$. All that can be said from a single configuration is that kicks larger
than $345\,{\rm km/s}$ are possible. Without more studies of the dependence of $v_{\rm kick}$ in $\phi$, we cannot in principle exclude the possibility
that kicks for these configurations could be nearly as large as in binary black hole mergers, although maximal values closer to those 
presented here appear more likely.

\section{Conclusions}

We continue our study of BHNS mergers at mass ratio $q=7$, the regime currently deemed to be the most likely for BHNS binaries in the field.
Previous results~\cite{2012PhRvD..85d4015F} have shown that for such mass ratios, high black hole spins $\chi_{\rm BH}>0.7$ are needed for the neutron star to disrupt, even
for large neutron stars ($R_{\rm NS}=14.4\,{\rm km}$). In this work, we look into the influence of the radius of the neutron star and the orientation of the black 
hole spin on the merger, focusing on configurations with mass ratio $q=7$ and black hole spin $\chi_{\rm BH}=0.9$.
We show that the transition between configurations for which the disruption of the neutron star causes the formation of massive accretion disks
and those where the neutron star just plunges into the black hole is very sensitive
to the compactness of the neutron star: while a $1.4M_\odot$ neutron star of radius $R_{\rm NS}=14.4\,{\rm km}$ leads to the formation of a massive disk ($M_{\rm disk}=0.2M_\odot$) 
and tidal tail ($M_{\rm tail}=0.25M_\odot$), a smaller neutron star $R_{\rm NS}=12.2\,{\rm km}$ in an otherwise identical binary forms a much less massive remnant 
($M_{\rm disk}=0.06M_\odot$, $M_{\rm tail}=0.09M_\odot$). For neutron stars with radii $R_{\rm NS}<10.5\,{\rm km}$, we expect no disruption at all. This indicates that
in the range of stellar radii currently favored~\cite{2010ApJ...722...33S}, a black hole spin $\chi_{\rm BH}\sim 0.9$ is required for a $1.4M_\odot$ neutron star to disrupt 
--- and thus for post-merger electromagnetic counterparts such as SGRBs and kilonovae to be possible. We also note that for all but the 
most massive disks a fairly low maximum density $\rho\sim 10^{11}\,{\rm g/cm^3}$ is observed, about an
order of magnitude lower than for disks of similar masses at lower mass ratios $q=3-5$. This is simply
the expected geometrical effect: the radius of the disk is roughly proportional to the mass of the final black hole. But that difference 
could significantly affect the late time evolution of the disk: the opacity of the disk to neutrino radiation will be lower, and the evolution of the magnetic
field is likely to be affected as well.

The amount of unbound material, approximated through measurements of the energy of fluid elements in the limit of a time-independent metric, is found to
be larger for these high-spin configurations than in previous, lower spin studies of BHNS mergers. The accuracy of these mass measurements is only
$\sim 50\%$ in our general relativistic simulations --- but for all three equations of state studied here, we find ejected mass $M_{ej}\go 0.01M_\odot$. 
The ejecta has a velocity distribution peaking at $v/c\sim 0.2$ (except for the larger neutron star, for which we find larger velocities which cannot be
accurately measured at this point) and a kinetic energy $E_{\rm kin} \go 10^{51}\,{\rm ergs}$. It would be a promising setup for a potential ``kilonova'', and could
even be detected as a radio afterglow. It should however be emphasized that these massive
ejecta are only produced in the high spin region of the parameter space, as the neutron star does not disrupt for low black hole spins.

The effect of a misaligned black hole spin is also studied in more details. General relativistic simulations of precessing BHNS binaries had only been
performed for one set of binary parameters before this work, for low mass, low spin black holes ($q=3$, $\chi_{\rm BH}=0.5$~\cite{2011PhRvD..83b4005F}). These high
spin configurations allow us to observe the effect of the misalignment of the black hole spin on tidal disruption more accurately. In particular, we
confirm that using the radius of the innermost stable spherical orbit $r_{\rm ISSO}(\chi_{\rm BH},\theta)$~\cite{2001PhRvD..64f4004H} as a way to predict the 
mass remaining outside the black hole at late times is reasonable: it matches the results of an aligned configuration with effective aligned spin $\chi_{\rm eff}$ and 
$r_{\rm ISCO}(\chi_{\rm eff})=r_{\rm ISSO}(\chi_{\rm BH},\theta)$~\cite{2012arXiv1209.4097S}. 
The qualitative features of the remnant can however be quite different, with the inclined
configurations being slower to form a disk, and keeping more mass in their tidal tail.

By extracting the gravitational-wave signal emitted by each of these BHNS mergers, we confirm that the effects of tides on the waveform of $q=7$ binaries
are negligible during most of the inspiral: tidal effects are below the numerical error in the simulation up to $f\sim 0.5\,{\rm kHz}$, and 
of the same order as the expected PN corrections~\cite{2010PhRvD..81l3016H,2012PhRvD..85l3007D} at higher frequencies.
The cutoff in the gravitational-wave spectrum at the frequency at which the neutron star disrupts is a slightly more promising imprint of the neutron star 
equation of state on the waveform. By combining PN results at low frequencies with our numerical waveforms, we estimate
that differences in the neutron star radius of order $2\,{\rm km}$ could be measured in at most $\sim 3\%$ of the Advanced LIGO events, for a single
detector at the current design sensitivity. Improved results can be obtained for the lower power lasers that are expected to be in use when Advanced LIGO
first begins to take data, if the detector is tuned to observe at $\sim 1.5\,{\rm kHz}$.
These estimates are however very optimistic, as they neglect the degeneracy between the neutron star radius and other binary parameters.

Finally, we observe that some precessing BHNS binaries receive significant kicks from the merger, as opposed to what was observed for non-precessing systems.
This is in agreement with results for BBH systems: the largest kicks are found for systems with partially misaligned black hole spins. In BHNS binaries, however, these
results are modified by the fact that after the neutron star disrupts, the gravitational-wave signal becomes weaker. As kicks mostly arise from the emission
of gravitational waves right around merger, their magnitude is significantly reduced for disrupting BHNS binaries. Accordingly, the largest kicks in BHNS systems
are found for binaries with misaligned spins for which the neutron star does not disrupt, or disrupts very late. Thus, the kicks measured at a mass ratio $q=7$ are
actually larger than for more symmetric systems, in opposition to BBH results. We find $v_{\rm kick}\sim 345\,{\rm km}$ for our most precessing system, which
represents a lower bound on the maximum kick attainable by the ensemble of all similar configurations with different orbital phases at merger.

One of the most important limitations of this work is that we do not model some critical physical effects: more realistic equations of state are required to study in
detail the evolution of the tidal tail and the characteristics of the ejecta, and are also a prerequisite for the inclusion of neutrino radiation. Neutrinos are the main source
of cooling in the disk, and cannot be neglected if we want to continue our evolution for longer than the few milliseconds after merger presented here. Finally, magnetic
fields should also play a crucial role in the evolution of the accretion disk --- although their evolution requires the use of a numerical grid much finer than what we currently
use in our simulations~\cite{2012PhRvD..86h4026E}. These effects are not expected to significantly affect the results presented here, which focused on the general 
properties of the merger and on the gravitational-wave signal, but they should be included in simulations aiming at a more detailed description of the evolution of 
BHNS systems after merger.

\acknowledgments

The authors wish to thank Kipp Cannon, Michael Boyle and Tanja Hinderer
for useful
discussions and suggestions over the course of this project,
Nicolas Smith-Lefebvre for computing and providing LIGO noise curves
tuned to high frequency,
the members of the SXS collaboration for their regular
suggestions, and the participants of the ``Rattle and Shine''
workshop at KITP (Santa Barbara) for stimulating discussions relevant to this work.
We thank Patrick Fraser for assistance with generating
Figs.~\ref{fig:gridinsp} and~\ref{fig:gridmerger} and John Wendell for
assistance with the other figures.  M.D. acknowledges support through
NASA Grant No.\ NNX11AC37G and NSF Grant PHY-1068243.  H.P. gratefully
acknowledges support from the NSERC of Canada, from the Canada
Research Chairs Program, and from the Canadian Institute for Advanced
Research.  L.K. and S.T. gratefully acknowledge support from the
Sherman Fairchild Foundation, and from NSF grants PHY-0969111 and
PHY-1005426.  C.D.O., M.S., and B.S. are partially supported by NASA
ATP grant no.\ NNX11AC37G and NSF grants PHY-1151197, PHY-1068881,
and PHY-1005655, by the Sherman Fairchild
Foundation, and the Alfred P. Sloan Foundation. F.F., L.K. and M.S.
were partially supported by the National Science Foundation
under Grant No.NSF PHY11-25915. Computations were
performed on the GPC supercomputer at the SciNet HPC
Consortium~\cite{scinet} funded by the Canada Foundation for
Innovation, the Government of Ontario, Ontario Research Fund--Research
Excellence, and the University of Toronto; on Briar{\'e}e from
University of Montreal, under the administration of Calcul Québec and
Compute Canada, supported by Canadian Foundation for Innovation (CFI),
Natural Sciences and Engineering Research Council of Canada (NSERC),
NanoQu{\'e}bec, RMGA and the Fonds de recherche du Qu{\'e}bec - Nature
et technologies (FRQ-NT); and on the Zwicky cluster at Caltech,
supported by the Sherman Fairchild Foundation and by NSF award
PHY-0960291.  This work also used the Extreme Science and Engineering
Discovery Environment (XSEDE), which is supported by National Science
Foundation grant number OCI-1053575.

\bibliography{References/References}

\appendix

\section{Summary of evolution equations}
\label{app:eqns}

The numerical simulations presented in this paper are evolved using the SpEC code~\cite{SpEC}, which allows us to solve
the coupled system formed by Einstein's equations of general relativity 
and the relativistic hydrodynamics equations. The SpEC code uses the
two-grid method~\cite{Duez:2008rb}, in which Einstein's equations are evolved using pseudospectral methods, while
the hydrodynamics equations are evolved on a separate finite difference grid. In this section, we first summarize the methods
used to evolve each system of equations independently, before discussing the communication between the two grids.

\subsection{Evolution of the metric}

The SpEC code uses the generalized harmonic formulation of Einstein's equations~\cite{Lindblom:2006}. 
The coordinates $x^b$ are assumed to obey the inhomogeneous wave equation
\beq
g_{ab}\nabla^c \nabla_c x^b = H_a(x,g_{ab})
\eeq
for an arbitrary function $H_a(x,g_{ab})$.
Einstein's equations can then be reduced to a set of symmetric hyperbolic first order equations for the metric 
$g_{ab}$, its spatial derivative $\Phi_{iab}\equiv \partial_i g_{ab}$, and its time derivative 
$\Pi_{ab}\equiv -t^c\partial_c g_{ab}$.
The principal part of the generalized harmonic equations (which we will denote by the symbol `$\simeq$') is
\beqn
\label{eq:evg}
\partial_t g_{ab} &\simeq& \beta^k \partial_k g_{ab}\\
\label{eq:evPi}
\partial_t \Pi_{ab} &\simeq& \beta^k \partial_k \Pi_{ab} -\alpha g^{ki} \partial_k \Phi_{iab}\\
\label{eq:evPhi}
\partial_t \Phi_{iab} &\simeq& \beta^k \partial_k \Phi_{iab} - \alpha \partial_i \Pi_{ab}
\eeqn 
which are equivalent to Einstein's equations as long as the constraints
\beqn
C_a &\equiv& H_a(x,g_{ab}) - g_{ab}\nabla^c \nabla_c x^b = 0 \\
C_{iab} &\equiv&  \partial_i \Psi_{ab} - \Phi_{iab} = 0
\eeqn
are satisfied (note that the standard Hamiltonian and momentum constraints are automatically satisfied if
$C_a = \partial_t C_a = 0$). Mathematically, satisfying the constraints in the initial conditions guarantees that
they will remain satisfied over the entire evolution. Small numerical errors in the evolution can however lead to
the exponential growth of constraint violating modes. To avoid such growth, a damping of the constraints is added
to the generalized harmonic system: we add $\gamma_1 \beta^i C_{iab}$ to equation (\ref{eq:evg}), 
$\gamma_0 \alpha (\delta^c_{(a} t_{b)}-g_{ab}t^c)C_c + \gamma_3 \beta^i C_{iab}$ to equation (\ref{eq:evPi}), 
and $\gamma_2 \alpha C_{iab}$ to equation (\ref{eq:evPhi}).
Choosing $\gamma_3 = \gamma_1\gamma_2$ guarantees that the system remains symmetric hyperbolic. To damp the constraints, we also
require $\gamma_0>0$ and $\gamma_2>0$. However, the values of $(\gamma_0$, $\gamma_2)$ that guarantee constraint
damping are not known analytically for an arbitrary metric. Choosing those damping parameters is thus largely
a trial-and-error process, whose success is gauged by verifying that the constraint-violating modes do
not grow significantly during the evolution, and that they converge to zero as the numerical resolution is
increased. In practice, we use, before disruption of the neutron star,
\beqn
\gamma_0 &=& 0.01 + \frac{4}{M_{\rm BH}} f(r_{\rm BH},w_{\rm BH}) + \nonumber\\
&& \frac{0.1}{M_{\rm NS}} f(r_{\rm NS},w_{\rm NS}) + \frac{0.2}{M}f(r_c,w_c),\nonumber\\
\gamma_2 &=& 0.01 + \frac{4}{M_{\rm BH}} f(r_{\rm BH},w_{\rm BH}) + \nonumber\\
&&\frac{1.5}{M_{\rm NS}} f(r_{\rm NS},w_{\rm NS}) 
+ \frac{0.6}{M}f(r_c,w_c),\nonumber\\
f(r,w)&=& e^{-r^2/w^2},
\eeqn
with $r_{\rm NS,BH,c}$ being, respectively, the coordinate distances to the center of the neutron star, the center of the black hole,
and the center of mass of the binary, while the widths are $w_{\rm BH}=2M_{\rm BH}$, $w_{\rm NS}=6M_{\rm NS}$ and $w_c=20 M$.
After disruption of the neutron star, we use
\beqn
\gamma_0 &=& 0.01 + \frac{12}{M_{\rm BH}} f(r_{\rm BH},w_{\rm BH}) + \frac{1.5}{M}f(r_c,w_c),\nonumber\\
\gamma_2 &=& \gamma_0,\nonumber
\eeqn
with now $w_{\rm BH}=3M_{\rm BH}$ and $w_c=20 M$. Finally, the parameter $\gamma_1$ is set to
\beq
\gamma_1 = 0.999 \left(f(r_c,10d) - 1\right) 
\eeq
with $d$ the coordinate separation between the black hole and the neutron star (this value is preferred to the previous choice
of $\gamma_1=-1$, which caused some characteristics speeds of the hyperbolic system to exactly vanish).

The gauge function $H_a$ is also freely specifiable in the generalized harmonic formalism.
We choose the initial value of $H_a$ by assuming that the time derivatives of the lapse $\alpha$ and shift $\beta^k$ vanish
in the coordinate frame corotating with the binary. During the inspiral, $H^a$ is evolved assuming that $t^aH_a$ and $H_i$ are
constant in the corotating frame. During merger, we follow the damped wave gauge prescription proposed by Szilagyi
et al.~\cite{Szilagyi:2009qz}, driving $H_a$ to
\beq
H_a = \mu_L \frac{\sqrt{g}}{\alpha}t_a - \mu_S \frac{g_{ai}\beta^i}{\alpha},
\eeq 
where $g$ is the determinant of the spatial metric $g_{ij}$, and 
we choose $\mu_L = 0$ and $\mu_S = \left[\log(\sqrt{g}/\alpha)\right]^2$ (Note that Ref.~\cite{Szilagyi:2009qz} recommends
$\mu_L = \left[\log(\sqrt{g}/\alpha)\right]^2$ instead. $\mu_L = 0$ was chosen because the evolution of the coordinates for
the collapse of isolated neutron stars was found to be better behaved than for $\mu_L = \mu_S$. However, in BHNS mergers, the
two choices appear to perform equally well, and future simulations will use $\mu_L = \mu_S$).
We smoothly transition between the two prescriptions, setting
\beq
\partial_t H_a = f(t) (\partial_tH_a)_{\rm damped} + (1-f(t)) (\partial_tH_a)_{\rm frozen},
\eeq
where $f(t)=1-e^{-(t-t_d)^4/w_d^4}$,  $t_d$ is the time at which
the damped wave gauge condition is turned on (typically, when matter starts accreting onto the black hole), and
$w_d=20M$ is the timescale over which the new gauge condition is turned on. $(\partial_tH_a)_{\rm damped,frozen}$
are the time derivatives of $H_a$ in, respectively, the `frozen' gauge used during inspiral, and the `damped wave' gauge
using during merger.

Another important feature of the code is the use of a time-dependent map between the coordinates of the numerical
grid and the inertial frame~\cite{2006PhRvD..74j4006S,2012arXiv1211.6079H}, which allows the grid to follow the orbital evolution of the
binary, and keeps the apparent horizon of the black hole spherical in the grid frame. The latter feature is required
because we excise a spherical region inside the apparent horizon of the black hole, and need all characteristic velocities
of the evolution equations to point out of the computational domain in order to avoid having to impose unknown boundary
condition on the excision surface. In practice, the map between the grid and inertial coordinates is the composition
of a distortion of the region immediately around the black hole of the type 
$r \rightarrow r + f(r) \sum_{lm} c_{lm}(t) Y^{lm}(\theta,\phi)$, which controls the size and shape of the excision surface 
($Y^{lm}(\theta,\phi)$ are the spherical harmonics functions, and $r$ the 
distance from the center of the apparent horizon),
a translation keeping the black hole center in place, and a rotation and global scaling to follow the orbital evolution of the binary.

The numerical grid is decomposed into a set of touching but non-overlapping subdomains, which are distorted cubes, spheres, balls and
cylinders (see Figs.~\ref{fig:gridinsp}-\ref{fig:gridmerger}). Boundary conditions between touching subdomains are treated using
a penalty method~\cite{Hesthaven1997,Hesthaven1999,Hesthaven2000,Gottlieb2001}, while the outer boundary uses an outgoing 
wave condition~\cite{Lindblom:2006}. As mentioned above no boundary condition is required on the excision surface.
In each subdomain, the solution is expanded on a set of basis functions dependent on the topology used (Chebyshev polynomials
for I1, Fourier basis for S1, spherical harmonics for S2 and Matsushima-Marcus functions~\cite{Matsushima:1995} for B3). For stable evolution, we filter
the evolved variables at the end of each time step by zeroing the top $N$ modes for each set of basis functions, with $N=1$ for
the I1 topology, $N=2$ for S1, and $N=4$ for S2 and B3 
(in which case the `top 4 modes' refers to modes with $l>l_{\rm max}-4$ in the decomposition
into spherical harmonics $Y_{lm}$). We should note that for the I1 topology, this is different from what was done in previous
BHNS simulations using the SpEC code. A stronger filter was then used, with 
\beq
a_n^{\rm filtered} = a_n \min{(e^{-\left(\frac{n/N-0.4}{0.4}\right)^6},1)},
\eeq
where $a_n$ is the $n^{th}$ coefficient of the spectral expansion, and $N$ is the total number of polynomials used in this set.
The filtering was also applied to the time derivative of the evolution variables, instead of to the variables themselves.
The weaker filtering used in recent simulations reduces the number of basis 
functions required to reach a given accuracy (a larger fraction
of the modes are unfiltered). 

Finally, the numerical resolution of the grid is chosen adaptively, and updated at regular intervals during the simulation.
The choice to increase or decrease the number of basis functions in any given set is done by comparing the truncation error
of the spectral expansion with a given target. We increase the resolution if the truncation error is below that target, and
decrease it if the truncation error after removing the top unfiltered mode would remain above it.

\subsection{Evolution of the fluid}

The neutron star is described as an ideal fluid with stress-energy tensor
\beq
T_{\mu \nu} = \rho_0 h u_\mu u_\nu + P g_{\mu \nu},
\eeq
where $\rho_0$ is the rest mass density of the fluid, $h$ the specific enthalpy, $P$ the pressure,
and $u^{\mu}$ the 4-velocity.
The general relativistic equations of hydrodynamics are evolved in conservative form, that is we evolve the `conservative' 
variables
\beqn
\rho_* &=& -\sqrt{g} n_\mu u^\mu \rho_0, \\
\tau &=& \sqrt{g} n_\mu n_\nu T^{\mu \nu} - \rho_*,\\
S_k &=& -\sqrt{g} n_\mu T^\mu{}_k,
\eeqn
where, as in the previous section, $g$ is the determinant of the spatial metric $g_{ij}$, while $n^{\mu}$ 
is the future directed unit normal to the time slice. 
Baryon number conservation
and the Bianchi identity $\nabla_{\mu}T^{\mu \nu}=0$ then allow us to write the evolution equations (see e.g.~\cite{2005PhRvD..72b4028D})
\begin{gather}
\partial_t \rho_* + \partial_j(\rho_* v^j) = 0,\\
\partial_t \tau + \partial_i(\alpha^2 \sqrt{g}T^{0i}-\rho_* v^i) = -\alpha \sqrt{g} T^{\mu \nu} \nabla_\nu n_\mu,\\
\partial_t S_i + \partial_j(\alpha \sqrt{g}T^j{}_{i}) = \frac{1}{2} \alpha \sqrt{g}T^{\mu \nu}\partial_i g_{\mu \nu},
\end{gather}
with $v^i$ the 3-velocity of the fluid. All evolution equations are in the conservative form
\beq
\partial_t u + \partial_i F^i = \sigma
\eeq
for some flux functions $F^i$ and source terms $\sigma$. To discretize these
equations on a finite difference grid, we need to compute the value of the fluxes $F$ at the interface between
numerical cells, and the source terms $\sigma$ at the center of each cell. A conservative scheme is mainly defined
by the method used to compute the fluxes. In SpEC, we use high
order shock capturing methods (WENO5~\cite{Liu1994200,Jiang1996202}) to reconstruct the physical
variables $\rho_0$, $T$ (as defined by Eqns~(\ref{eq:eosP}-\ref{eq:eosE})) and $u_i$ at cell faces from their values at cell centers.
The WENO5 algorithm gives us for each reconstructed variable $v$ and on each face a left state $v_L$ and a right state $v_R$.
Both $v_L$ and $v_R$ are computed using a five-point stencil, with $v_L$ using three points on the left of the face, and $v_R$ only two.
In smooth regions, both $v_L$ and $v_R$ are fifth order accurate interpolations of $v$ on the face. In the presence of a shock,
the reconstruction is only first order accurate, and attempts to reconstruct $v$ using a 3-point stencil which does not
include the location of the shock. From these reconstructed variables, we can then compute the fluxes $F_L$ and $F_R$. The approximate
Riemann problem on each face is then solved by computing the HLL flux~\cite{HLL}
\beq
F = \frac{c_{\rm min} F_R + c_{\rm max} F_L - c_{\rm min} c_{\rm max} (u_R-u_L)}{c_{\rm max}+c_{\rm min}},
\eeq
where $(c_{\rm min},c_{\rm max})$ are the left-going and right-going characteristic speeds.

To evolve these equations on a finite difference grid, two additional modifications are required. The first is a correction to low-density
regions, where small numerical errors in the evolved (`conservative') variables can lead to large or unphysical values for the physical
variables ($T$,$h$), or negative values of the density. That correction is applied in regions in which 
$\rho_0<10^{-6}\rho_0^{\rm max}(t)$, where $\rho_0^{\rm max}(t)$ is the maximum value of the rest mass density at the current time.
In that region, we set $T=0$ and $u_i=0$ (more precisely, we do so when $\rho_0<10^{-7}\rho_0^{\rm max}(t)$, and apply
ceilings $T<T_{\rm max}$ and $g^{ij}u_iu_j<u^2_{\rm max}$ in the intermediate region $10^{-7}<\rho_0/\rho_0^{\rm max}<10^{-6}$, where $T_{\rm max}$
and $u_{\rm max}$ are linear functions of $\rho_0$). We also require
$\rho_0>10^{-11}\rho_0^{\rm max}[t=0]$ to avoid negative densities. The second correction occurs when the `conservative' variables do not
correspond to any set of physical variables. This occurs when
\beq
\frac{S^iS_i}{\rho_*^2} > \tilde{S}^2_{\rm max}\equiv \frac{\tau}{\rho_*}\left(2+\frac{\tau}{\rho_*}\right).
\eeq
To avoid this, we impose
\beq
\frac{S^iS_i}{\rho_*^2} \leq \alpha \tilde{S}^2_{\rm max}
\eeq
at all times, with $\alpha = 0.999$ for $\rho_*\geq 10^{-3}\sqrt{g}\rho_o^{\rm max}$, and 
\beq
\alpha=0.999-0.0005 \log_{10}{\frac{\rho_*}{10^{-3}\sqrt{g}\rho_o^{\rm max}}}
\eeq
at lower densities. 

We should also note that the finite difference grid on which we evolve the fluid variables is modified at regular intervals during the
evolution. Indeed, the main advantage of the two-grid method is that the finite difference grid only needs to cover the region
in which matter is present. This region does, however, vary over the course of the evolution. To automatically adapt the grid
to the current location of the fluid, we measure the flow of matter across two surfaces, located $\sim 0.05L$ and $\sim 0.2L$ away
from the outer boundary (where $L$ is the size of the grid). When a significant flow of matter crosses the outer surface, we expand
the grid in the direction in which an outflow is detected (`significant outflow' is here defined
as a flow of matter sufficient to lose $\sim 0.1\%$ of the neutron star mass if it was maintained at this level for the entire
evolution). Similarly, we contract the grid when matter is no longer detected at the inner surface. The expansion of the grid
is limited to $\sim 20M_{\rm BH}-50M_{\rm BH}$, and after disruption most of the grid points are focused in the region in which the 
accretion
disk forms. More details on the exact map used between the finite difference grid and the spectral grid can be found in
Ref.~\cite{2011PhRvD..83b4005F}.

\subsection{Time-stepping and source term communication}

Both sets of equations are evolved jointly, using the same time-stepping 
method: we use third-order Runge-Kutta (RK3) time-stepping, with adaptive choice
of the time step. The time step is chosen by comparing the results of the RK3 algorithm with a second-order method that does not require any 
new computation of the time derivatives of the variables, as described in Chapter 17.12 of Ref.~\cite{numrec_cpp}. At our medium resolution
and during inspiral,
we require the relative error in the evolution of the variable at any grid point to be smaller than $\epsilon_{\rm rel}=10^{-4}$, or the
absolute error to be smaller than $\epsilon_{\rm abs}=10^{-6}$ (or, for fluid variables which naturally scale like $\rho_0^{\rm max}(t=0)$, 
$\epsilon_{\rm abs}=10^{-6}\rho_0^{\rm max}(t=0)$). During mergers, we multiply these values by a factor of $10$. At different resolutions,
the error is chosen to scale like the second-order error in the finite difference evolution (i.e., for finite difference
grids with respectively $N_1$ and $N_2$ grid points along each dimension, we choose $\epsilon(N_1)*N_1^2=\epsilon(N_2)*N_2^2$).

The last important part of our evolution algorithm is the communication between the spectral grid on which we evolve the metric and the
finite difference grid on which we evolve the fluid variables. At the end of each time step, we interpolate from the spectral grid
onto the finite difference grid the metric quantities and their derivatives in the coordinates
of the numerical grid which are required for the computation of the right-hand side of the hydrodynamics equations 
($g_{ij}$, $K_{ij}$, $\alpha$, $\beta^i$,$\partial_ig_{jk}$, $\partial_i\alpha$, $\partial_i\beta^j$). Similarly, we interpolate from the
finite difference grid to the spectral grid the fluid variables required to compute the stress-energy tensor for the evolution
of Einstein's equations ($\rho_0$, $h$, $W=\alpha u^0$, $P$, $u_i$). Because spectral interpolation of the metric variables onto
every point of the finite difference grid would be extremely expensive, the first interpolation is done by refining the spectral
grid by a factor of 3 (a cheap operation, as it solely requires adding basis functions whose coefficients are all zero), and then
using that refined grid to perform fourth-order accurate polynomial interpolation onto the finite difference grid. Interpolation
from the finite difference grid onto the spectral grid uses third-order shock capturing interpolation (WENO3).
Finally, we should note that grid-to-grid interpolation is only performed at the end of each time step, and not at the intermediate
time steps taken by the RK3 algorithm. To obtain the values of the interpolated variables at intermediate time steps, we extrapolate in
time from their values at the last two interpolation times.

\section{Combining PN and numerical results to assess the detectability of Equation of State effects in the gravitational waveforms}
\label{app:match} 

The influence of tidal effects in the low-frequency part of the waveform can
be estimated from simple PN considerations:
to leading order, the amplitude of the Fourier transform of the waveform is
\beq
A(f) = \frac{M^{5/6}}{D\pi^{2/3}} \sqrt{\frac{5\eta}{24}} f^{-7/6}
\eeq
while its phase is
\beq
\Psi(f) = \Psi_0(f) + \Psi_T(f)
\eeq
where $\Psi_T(f)$ contains the tidal effects, and $\Psi_0(f)$ all other contributions.
In the regime considered here, $\Psi_0(f)$ is poorly constrained. However, it is also
identical for all 3 of our non-precessing configurations. Thus, we can compute
the inner product $\|\delta h\|$ for two BHNS binaries which only differ by the
equation of state of their neutron star using $h_1(f)=A(f)$ and 
$h_2(f)=A(f)e^{i(\Delta\Psi_T^{\rm PN}(f)+\Delta\phi+af)}$, where $\Delta\Psi^{\rm PN}_T(f)$ is the phase difference due to
tidal effects (computed from Eq.~(\ref{eq:PsiPN})), $\Delta\phi$ is an arbitrary phase shift, and
$a=2\pi(\Delta t)$ allows for an arbitrary time shift. $a$ and $\Delta\phi$ are chosen 
to maximize $\langle h_1,h_2 \rangle$. Limiting our integration to frequencies below $0.8\,{\rm kHz}$,
we find that tidal effects during the inspiral are a slightly smaller effect
than the disruption of the neutron star for the Zero Detuned noise curves (see Table~\ref{tab:diff}).

Considering the low-frequency and high-frequency portions of the waveforms 
separately result in an underestimate of the
total $||\delta h||$. Indeed, each part of the waveform used different time and phase shifts to maximize
$\langle h_1,h_2 \rangle$. We would thus expect 
$||\delta h||^2_{\rm tot}>||\delta h||^2_{<0.8\,{\rm kHz}}+||\delta h||^2_{>0.8\,{\rm kHz}}$.
If we assume that the tidal part of the PN approximation is 
valid at the beginning of the simulation (which is
approximately true, as shown in Fig.~\ref{fig:gwphiomega}), and that the simulations only differ by
the effects of the neutron star equation of state (neglecting the residual eccentricity, as well as
numerical errors), we can however obtain reasonable estimates of the total $||\delta h||$.
Indeed, we can write the waveforms from two numerical simulations as
\beqn
\tilde h_1(f) &=& A_1(f) e^{i(\Psi_0(f)+\Psi_{T,1}(f))}\\
\tilde h_2(f) &=& A_2(f) e^{i(\Psi_0(f)+\Psi_{T,2}(f))}.
\eeqn
As for the PN expansion, the phase $\Psi_0(f)$ does not contribute to $||\delta h||$,
while $\Delta \Psi^{NR}_T=\Psi_{T,2}(f)-\Psi_{T,1}(f)$ can be matched to the Post-Newtonian $\Delta \Psi^{\rm PN}_T(f)$ over a given
frequency range. Practically, we compute $\Delta \Psi^{NR}_T$ from the simulations, allowing
for an arbitrary time and phase shift in one of the numerical simulations. The shifts are chosen to
minimize the difference with the PN predictions in the
range $0.3\,{\rm kHz}<f<0.8\,{\rm kHz}$ (modifying the matching window has a $\sim10\%$ effect on $||\delta h||$,
similar to the differences between the various PN orders). We then smoothly connect the amplitude and the phase
of the PN and numerical waveforms through $Y_{\rm tot}(f)=a(f)Y_{\rm PN}(f)+(1-a(f))Y_{\rm NR}(f)$
with
\beq
a(f) = 0.5*\left(1 + \cos\left(\frac{\pi (f-f_l)}{f_u-f_l}\right)\right)
\eeq
for $f_l<f<f_u$ [where $f_l$ and $f_u$ are the bounds of the matching window and
$Y(f)$ is either $A(f)$ or $\Delta \Psi_T(f)$]. We also have $a(f<f_l)=1$ and
$a(f>f_u)=0$. The resulting hybrids contain the information needed to estimate the difference $||\delta h||$
between waveforms. It is worth noting, however, that they are not proper hybrid waveforms, and would be 
useless as templates. Indeed, we used the fact that the ony difference between the 3 systems considered here is
the equation of state of the neutron star to neglect the non-tidal part of the
phase, $\Psi_0(f)$. However, knowledge of $\Psi_0(f)$ would be needed in order to compare an observed waveform
to the result of our simulations.

\end{document}